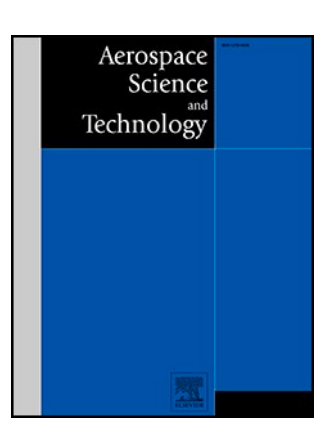

# Construction of flexible wing models by combined manufacturing of additive and subtractive processes for transonic wind tunnel testing

Natsuki Tsushima [a,b,c,*], Kensuke Soneda [c], Kenichi Saitoh [c], Kazuyuki Nakakita [c]

[a] Department of Aeronautics and Astronautics, Kyushu University Nishi Ward, Fukuoka 819-0395, Japan
[b] Department of Aeronautics and Astronautics, The University of Tokyo Hongo, Tokyo 131-8656, Japan
[c] Aviation Technology Directorate, Japan Aerospace Exploration Agency Mitaka, Tokyo 181-0015, Japan

ABSTRACT

Although numerical capabilities to evaluate the aeroelastic characteristics of aircraft have improved significantly over the past few decades, wind tunnel experiments continue to play a crucial role in aerospace research and development, as they are essential for predicting aircraft performance and validating numerical solutions. Previous works by the authors have proven the feasibility of additively manufactured wing models for transonic wind tunnel testing. This work extends prior studies by systematically combining additive manufacturing and subtractive machining processes for constructing flexible wing models for high-speed wind tunnel testing, thereby significantly enhancing manufacturing effectiveness and reproducibility. This systematic approach addresses the limitations of previous methods, such as those relying on skill-dependent mechanical polishing, by ensuring stable quality across multiple fabricated models and enabling repeatable test data. The potential of wing models fabricated by the present manufacturing method was first investigated by evaluating their geometrical accuracy. The structural characteristics of such wing models were then investigated by comparing them with finite element solutions. Finally, a transonic wind tunnel testing with fabricated wing models was performed. The fabricated wing models showed good geometrical accuracies and realizations of structural characteristics. These results also demonstrated that the combined manufacturing technique could fabricate flexible wind tunnel models with great reproducibility. Specifically, the average surface roughness was <1.0 μm, and the average surface deviation was <0.3 mm, showing an improvement in geometrical precision compared to previous methods. The flutter frequencies for Wing 1a and 1b were 157.0 Hz and 158.0 Hz, respectively, which agreed very well, highlighting the excellent reproducibility of aeroelastic behavior across different models. The present manufacturing technique would help to effectively produce wind tunnel models at a low cost with a short lead time. The good reproducibility of the present approach also enables it to obtain repeatable test data with different models.

## 1. Introduction

The aeroelastic characteristics of aircraft must be evaluated in the design of modern aircraft to ensure performance and safety. Responses of aircraft/wings in the transonic regime, in particular, involve various nonlinear phenomena like shock waves, which might risk the safety of aircraft with unstable responses. Although numerical capabilities to evaluate the aeroelastic characteristics of aircraft have been improved over the past decades [1–8], wind tunnel experiments remain to play an important role in aerospace research and development to investigate aircraft performance and validate numerical solutions. To precisely capture phenomena of interest, a wind tunnel model has to be constructed by accurately realizing the aeroelastic characteristics of the designed aircraft/wings. Also, aeroelastic scaled models are often designed for wind tunnel testing to estimate the aeroelastic responses of full-scale aircraft/wings [9–12]. The process of aeroelastic scaling involves a simultaneous consideration of structural and aerodynamic physics. Aerodynamic similarity can usually be achieved by geometrically scaling the outer mold line [13]. On the contrary, structural similarity cannot be realized with a simple scaling of the structural geometries as it may lead to unrealistic material properties for the scaled model. Moreover, the structural components at a smaller scale may result in difficulties in fabrication. As a consequence, the design and construction of wind tunnel models have traditionally required a highly skilled and time-consuming process [14]. Therefore, it has been desired to establish an efficient approach for model construction to realize high-quality models and improve data productivity from wind tunnel experiments while reducing cost and time for the model construction.

* Corresponding author.
*E-mail address:* ntsushima@aero.kyushu-u.ac.jp (N. Tsushima).

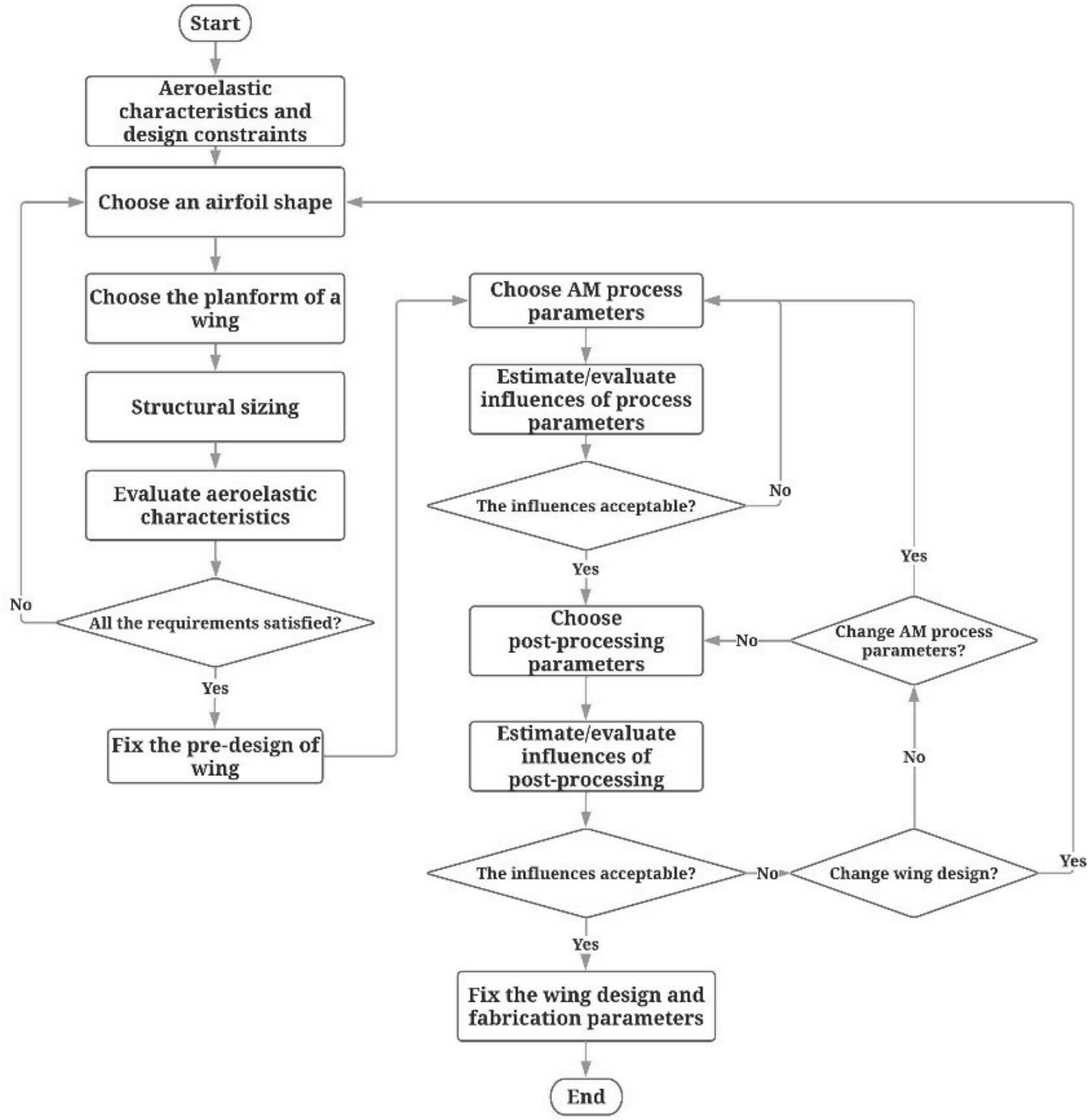

**Fig. 1.** An algorithm for the modeling and construction of wing models with combined manufacturing.

The recent advancement of additive manufacturing (AM) technology has enabled the realization of conceptual structures that had been difficult to fabricate [15–17]. The capabilities and potential of AM technology to develop various sophisticated structures have been demonstrated in the literature [18–21]. Applications of the AM technology for effective wind tunnel testing have also been explored extensively [22]. Additively manufactured models have shown the capability to produce reliable results in wind tunnel experiments for aerodynamic evaluations with simple experimental conditions in which models are considered “rigid” [23]. With the higher degrees of design and manufacturing freedom, the AM technique could extend the capability and enable us to establish an effective way for the construction of aeroelastic wind tunnel models. Recent achievements in AM have particularly facilitated the development of complex compliant structures for morphing applications, lightweight components through topology optimization for aerospace designs [24,25], and rapid prototyping of aeroelastic test articles for experimental characterization [26,27]. On the other hand, the properties of additively manufactured structures are strongly correlated to the printing process variables [28–30]. Therefore, the accurate construction of an aeroelastic wind tunnel model based on the AM technique requires precisely considering the influences of the process variables on the properties of additively manufactured structures.

High stiffness and strength properties are required in models for high-speed wind tunnel testing to withstand high aerodynamic loads. The Powder Bed Fusion (PBF), providing compatibility with metal alloys, is commonly used to achieve high stiffness and strength properties. However, metal AM-based models usually result in insufficient surface roughness, which is critical to obtaining reliable aerodynamic data during wind tunnel testing. Therefore, most AM-based models require after-treatments. Our previous works studied aeroelastic wing models for transonic wind tunnel testing based on the metal AM technique [31, 32]. Mechanical polishing (MP) and electric discharge machining (EDM) were applied for smooth surfaces and precise trailing edge shapes. While these AM-based wing models provided reliable aeroelastic experimental results, the prior construction method combined with the MP process highly depended on the skills of engineers, resulting in high variation rates in final models. Therefore, the method of wind tunnel model fabrication must have been further improved to ensure stable qualities in final models systematically.

**Table 1**
Geometries of wing models.

| Property | Wing 1 | Wing 2 |
|---|---|---|
| Semi-span, mm | 200.00 | 249.00 |
| Root chord, mm | 50.00 | 100.58 |
| Taper ratio at root (kink) | 1.00 | 0.23 (0.54) |
| Sweep angle (on leading edge), deg | 20.00 | 27.63 |
| Incident angle at root (tip), deg | 1.00 | 5.81 (1.22) |
| Airfoil profile | NACA0007 | Supercritical |

This paper proposes a novel methodology for a wing model construction by the combined manufacturing technique of metal additive and subtractive processes for transonic wind tunnel testing as an extension of the previous works by the authors. This approach significantly enhances reproducibility and provides stable quality compared to previous methods, enabling reliable and repeatable experimental results. The objectives of this paper are 1) to investigate the geometrical precisions of wing models constructed by the combined manufacturing technique, 2) to study the structural and aeroelastic characteristics of wing models fabricated by the present approach, and 3) to explore the feasibility and capability of the wing models for transonic wind tunnel experiments.

In the present approach, the wing models are first fabricated as “as-built” models with near-net shapes based on the PBF process, direct metal laser sintering (DMLS), by using a metal AM machine. Powders of an aluminum-based alloy (AlSi10Mg) are used for the fabrication. The subtractive machining (SM) process is then applied to the “as-built” models to realize better qualities of surface roughness and trailing edge shape. Two different wing designs are fabricated to investigate variations of the geometrical accuracy in different wing geometries. The geometrical accuracies of the constructed wing models are evaluated by using a 3D scanning technique. The static and dynamic structural characteristics of the wing models are also evaluated with static load tests and ground vibration tests (GVTs). The transonic aeroelastic characteristics of the fabricated wing models by the proposed technique are finally investigated through transonic wind tunnel testing. The characteristics of the AM-based wing models are also discussed.

## 2. Wing model

The feasibility and capability of the present combined manufacturing technique were evaluated by constructing wing models with the combined manufacturing processes. The process consisted of three sub-processes: initial design, AM process, and SM process. The AM process was conducted based on DMLS by using EOS M290 with AlSi10Mg powders. The SM process adapted a computer numerical control (CNC) mechanical cutting machining. The processes involved the initial modeling and parameter determinations of combined processes as shown in Fig.1. The present combined manufacturing, performing the additive and subtractive processes in series, allows us to estimate individual influences on models in structural properties and geometrical precision, which improves the controllability of each process. By following the determinations of model and manufacturing parameters, as-built wing models were first fabricated as near-net shapes with allowances for the SM process. The CNC mechanical machining process was conducted on the surfaces and trailing edges of wing models.

Two different port wing models were designed for this study. Table 1 summarizes the geometries of the wing models. The first model was a thin rectangular swept wing. The airfoil profile was a NACA0007. The second model was a more practical double-tapered swept wing with a supercritical airfoil. Figs. 2 and 3 show the planforms and cross-sections at the root of the wing models. The wing models were designed as hollow structures with thin upper and lower surfaces, whose aeroelastic characteristics could simply be controlled by a single parameter (i.e., the surface thickness). One advantage of the AM technique is that it allows adopting/fabricating such simple but effective designs to control the aeroelastic characteristics of wing models. The upper and lower surfaces of Wings 1 and 2 had thicknesses of 1.00 and 1.50 mm, respectively. The 5-mm/5.5-mm front sections from the leading edges were solid structures for each wing. The tip of the Wing 1 model was closed with a 1.00-mm wall. There was a small hole on the tip wall to remove residual metal particles inside the model after the AM process. The Wing 2 model had a solid region around the tip (see Fig. 3). Residual metal particles in the Wing 2 model were removed from the root base with small holes after the AM process. Solid base fixtures were integrated with the wing models for installation in a wind tunnel. The incident angle of the Wing 1 model was 1°, while the incident angles at the root and tip of the Wing 2 model were 5.81° and 1.22°, with a geometric twist. The swept wing models were fabricated by stacking layers from the base to the tip with additional supports for accurate realization of airfoil profiles, as described in Fig. 4. As-built models were first fabricated as near-net shapes with extra surface thicknesses to provide allowances for the following SM process. The extra surface thicknesses on the wing surfaces for the post-process allowance $\delta$ were about 0.5 mm. The trailing edges were also slightly extended for the machining process (i.e., about 3.5 mm for the Wing 1 model). The surfaces and trailing edges of the wing models were then finished by computer numerical control (CNC) mechanical cutting. In general, the required surface roughness is set to be $<$10 μm [33,34] (typically around 0.8 to 3.2 μm [22]) to obtain reliable aerodynamic characteristics in wind tunnel testing. Table 2 summarizes the fundamental printing process variables used for the AM process. The wing models were built by NTT Data XAM Technologies Corp. and post-processed by Hitachi Metals, Ltd. (Proterial, Ltd.). For the Wing 1 model, two models (Wings 1a and 1b) were constructed to evaluate the variations in different fabrications for the same design.

## 3. Geometrical evaluation of fabricated wing models

In this chapter, the geometrical precisions of the fabricated as-built and final wing models were evaluated. The surface deviations and

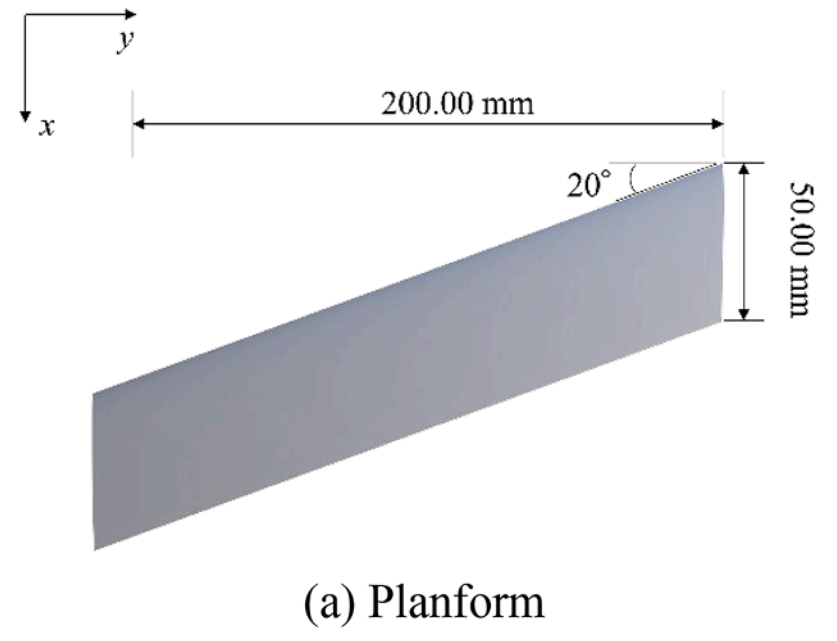

(a) Planform

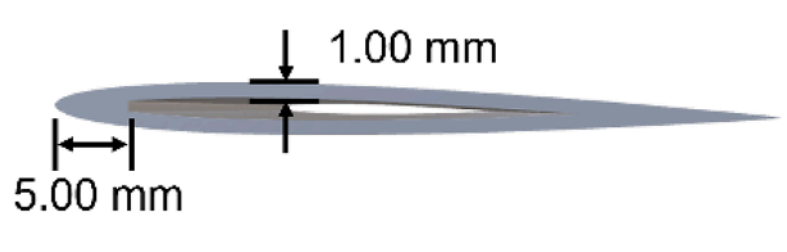

(b) Cross-section at the root

**Fig. 2.** The design of the Wing 1 model with a NACA0007 airfoil.

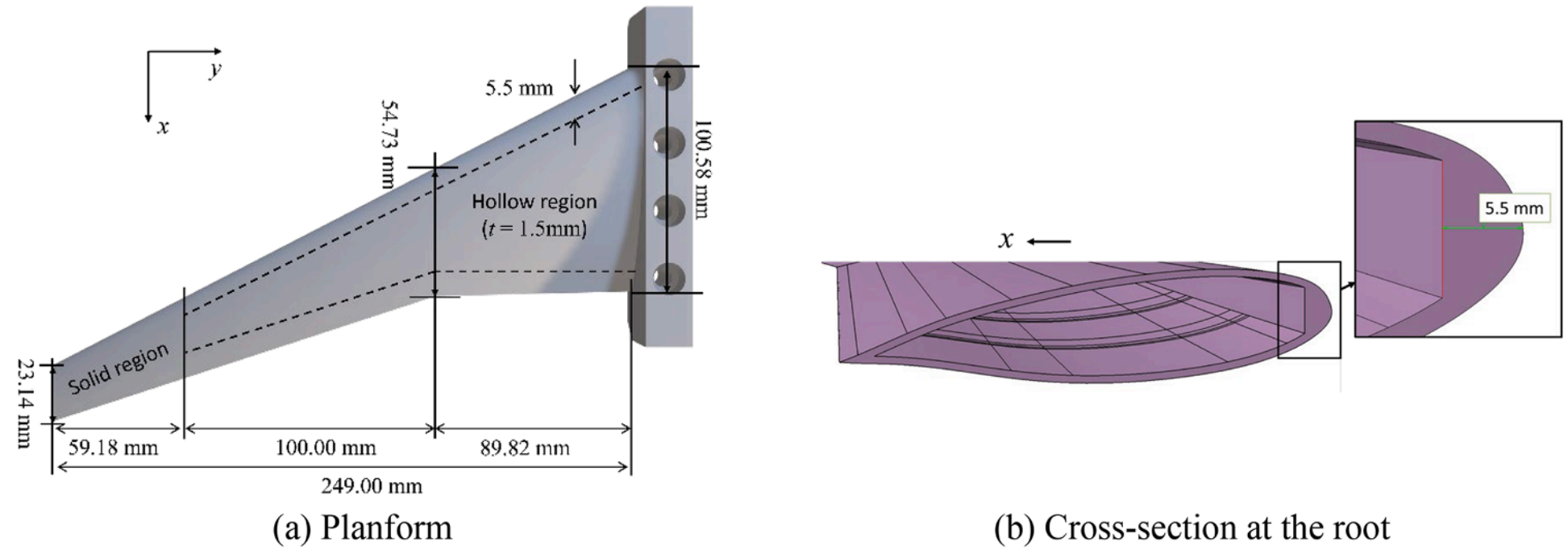

(a) Planform (b) Cross-section at the root

**Fig. 3.** The design of the Wing 2 model with a supercritical airfoil.

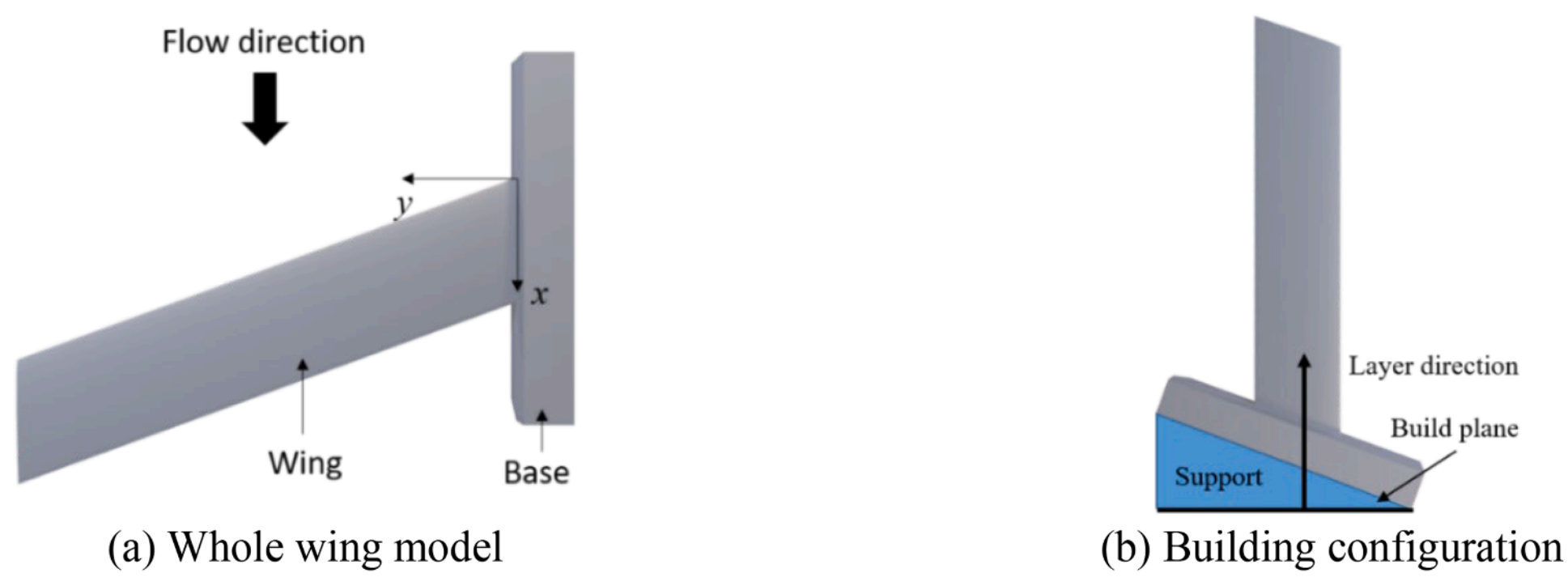

(a) Whole wing model (b) Building configuration

**Fig. 4.** The whole Wing 1 model and its building configuration.

**Table 2**
Fundamental process variables for the AM.

| Property | Value |
|---|---|
| Powder | AlSi10Mg |
| Layer height, μm | 60 |
| Build platform temperature, °C | 200 |

cross-sectional geometries of the fabricated models were measured by a 3D scanning technique. The internal geometries and surface roughnesses of the final models were also evaluated.

### 3.1. Geometrical accuracy of as-built models

As the present combined manufacturing involves the sequential additive and subtractive processes, the geometrical precisions of the as-built models were first investigated. The three-dimensional geometries of the as-built models were obtained by a 3D scanning system, ATOS Q 8 M (GOM). The surface deviations in the as-built models were evaluated by comparing the measured geometries of the as-built models with those of the original computer-aided design (CAD) models for the additive processes. The inspection of the surface deviations was performed by GOM Inspect (GOM).

Fig. 5a shows the geometrical comparison between the original CAD model and the corresponding as-built model for Wing 1a. The models in navy and gray represented the CAD and the corresponding as-built models. The small grip on the wing tip of the as-built model for the machining process was removed after the SM process. The models were aligned with three surfaces on the base fixture of the wing model for comparison purposes, simulating the fixed boundary conditions expected for wind tunnel testing. Fig. 5b represents the surface deviations in the $z$-direction of the as-built Wing 1a model along the spanwise direction at three chordwise locations (around the leading edge, mid-chord, and trailing edge) relative to the original CAD model. It should be noted that the as-built Wing 1b model showed almost identical results to Wing 1a, confirming high consistency in the AM process itself, even before the SM process. For Wing 1a, the deviations on the inner span were mostly $<0.03$ mm, with a maximum deviation near the wing tip of approximately 0.13 mm. The cross-sectional comparison at the tip (Fig. 6) shows a slight downward deformation in the as-built model, but its influence on the overall geometry was considered negligible as the deviations were localized to the tip. The twist of the as-built wing model along the span was $<0.05^\circ$ compared with the original CAD model. Therefore, it was confirmed that the as-built model of Wing 1 was manufactured with good precision for the near-net shape.

Fig. 7a shows the geometrical comparison between the original CAD and as-built models for Wing 2. The CAD and as-built models for Wing 2 are indicated in yellow and gray. Similar to Wing 1, the alignment method used for Wing 2 also adopted the same three surfaces on the base fixture, consistent with the wind tunnel test setup. Fig. 7b illustrates the surface deviations of the as-built Wing 2 model along the span at three chordwise locations. The inner span of the as-built Wing 2 model achieved good agreement with the original CAD model. However, due to its complex geometry (supercritical airfoil with sweep and taper), larger deviations were observed compared to Wing 1. A maximum deviation near the wing tip was approximately 0.32 mm. Even for the supercritical airfoil, the airfoil shape was accurately reproduced, although the as-built model was slightly deformed upward, as shown in the cross-sectional comparison at the tip (Fig. 8). The large deviation in the trailing edge could be disregarded as the edge was treated by the SM process. The twist of the as-built Wing 2 model along the span was $<0.08^\circ$ compared with the original CAD model. Therefore, the as-built model of Wing 2 was also manufactured with reasonable precision for the near-net shape, despite the larger deviations than Wing 1 due to the more intricate geometry.

### 3.2. Geometrical accuracy of the final models

The as-built wing models were further processed by CNC cutting for final shapes. The geometrical accuracies of the final wing models were

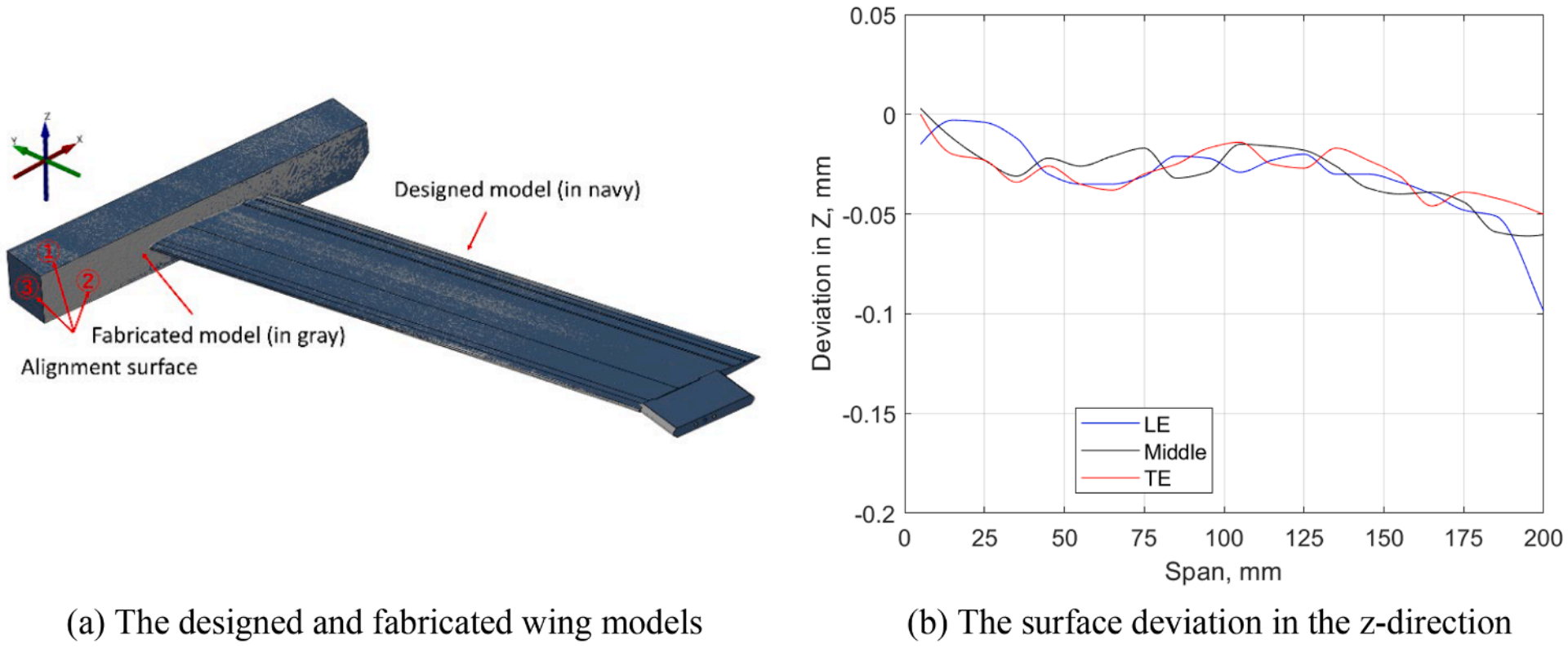

(a) The designed and fabricated wing models (b) The surface deviation in the z-direction

**Fig. 5.** The comparisons of the as-built Wing 1a models and the surface deviation in the z-direction along the span at three chordwise locations.

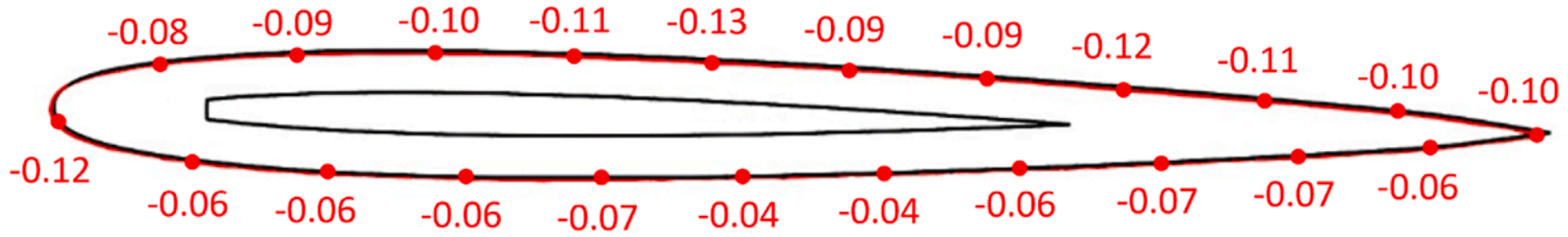

**Fig. 6.** The deviation of the cross-sectional geometry in the normal direction for the as-built Wing 1a model.

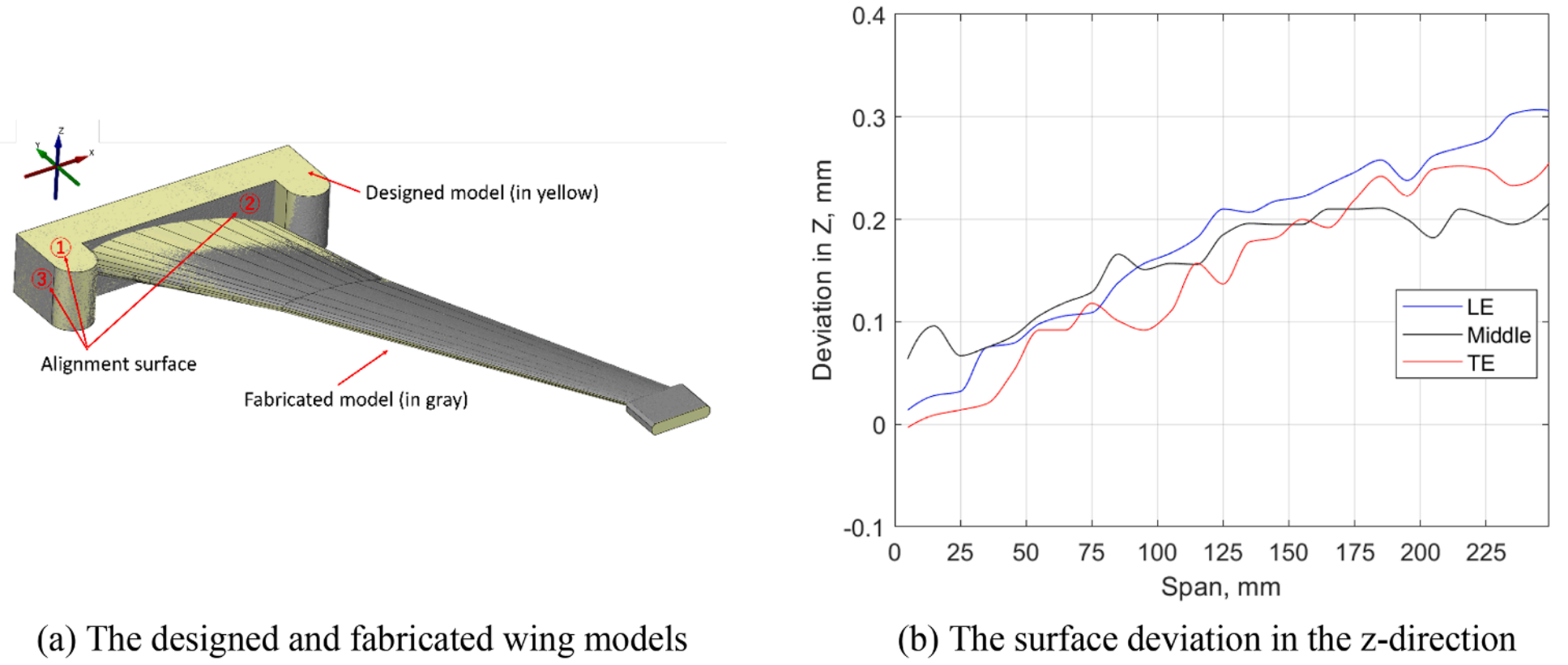

(a) The designed and fabricated wing models (b) The surface deviation in the z-direction

**Fig. 7.** The comparisons of the as-built Wing 2 models and the surface deviation in the z-direction along the span at three chordwise locations.

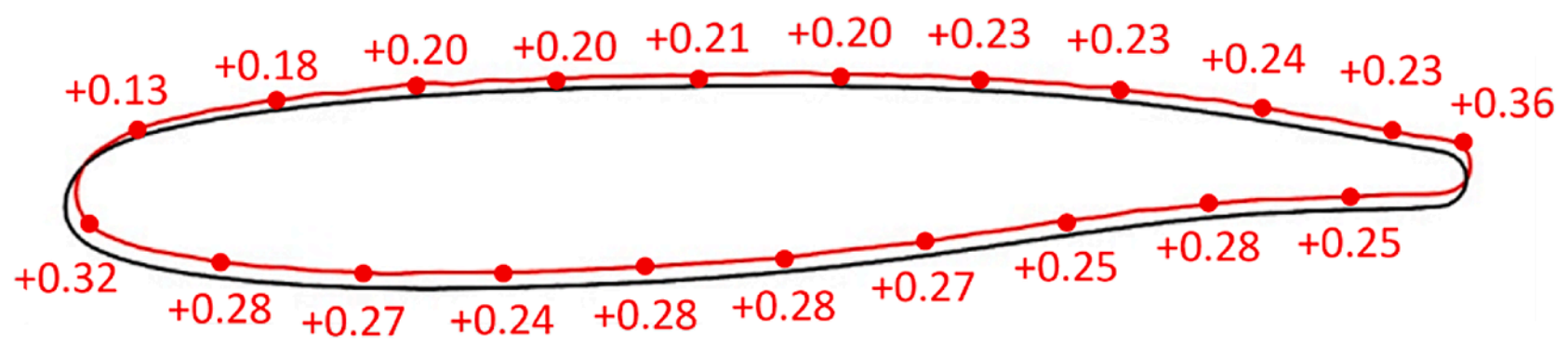

**Fig. 8.** The deviation of the cross-sectional geometry in the normal direction for the as-built Wing 2 model.

evaluated in this section. Fig. 9 shows the example picture of the final model for Wing 1a. The difference in the texture on the wing surface along the span was due to the difference in the movement of the cutting edge during the machining process.

Fig. 10a compares the geometries of Wing 1 between the original CAD model and the corresponding final model. The CAD and final models are represented by purple and gray, respectively. The models were aligned by using the same method as the previous cases. Fig. 10b presents the surface deviations of the final Wing 1a model along the span at three chordwise locations. As with the as-built models, the results for Wing 1b showed the same excellent consistency as Wing 1a, reinforcing the high reproducibility of the combined manufacturing process. The deviations in the $z$-direction on most inner surfaces were <0.2 mm. A maximum deviation of about 0.5 mm (a slight downward warping as shown in the cross-sectional comparison in Fig. 11) occurred around the tip. The warping in the fabricated final model was partially attributed to the deformation in the as-built model. As the machining process released some residual stress, which usually occurs in additively manufactured structures, the wing model tended to further induce warping. The average twist deviation of the final models along the span was <0.1° compared with the CAD model. According to our previous study [31], it was confirmed that the average surface deviations along the leading and trailing edges of wing models with similar configurations were about 0.3 mm. For the present approach, the average deviations along the leading and trailing edges were <0.3 mm. Therefore, the geometrical precision was improved by the present approach, while providing stable quality for manufacturing multiple wing models.

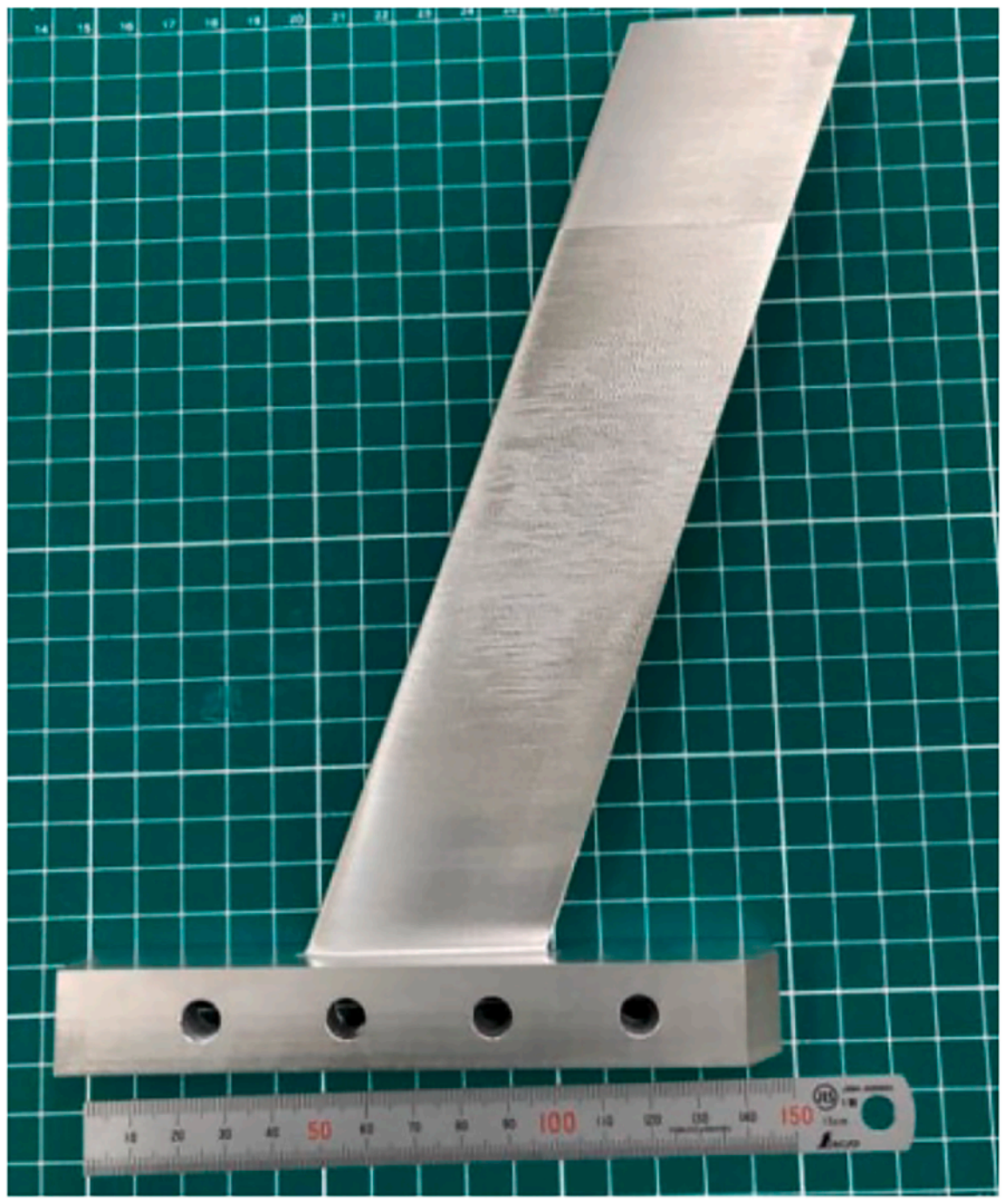

**Fig. 9.** A picture of the final model for the Wing 1a model.

Fig. 12 shows the picture of the final model for Wing 2. Fig. 13a compares the geometries of Wing 2 between the original CAD model (in navy) and the corresponding final model (in gray). Fig. 13b illustrates the surface deviations of the final Wing 2 model along the span at three chordwise locations. The comparisons of the cross-sections around the mid-spans and tips of the CAD and final models are also shown in Fig. 14. The maximum deviation of 0.15 mm in the $z$-direction occurred around the tip, while the rest of the wing surfaces showed a very good agreement with the designed model. The average surface deviation of the Wing 2 model in the $z$-direction was 0.03 mm. The average twist deviation of the final Wing 2 model along the span was also <0.1° Therefore, it was confirmed that the present manufacturing approach was capable of precisely realizing practical wing designs even with sweep, double-taper ratio, geometric twist, cambered airfoil profiles, etc.

The overall external geometries of the Wings 1 and 2 models are summarized in Table 3. For the final Wing 1 model, the values are the averages of Wings 1a and 1b The parentheses denote the percentage differences in the geometries of the CAD and fabricated models in the final designs. The differences in the semi-span of the final models to the corresponding CAD models were 0.04 and 0.08 mm, which were <0.1 % differences. The differences in the root and tip chords were 0.06 mm for the final Wing 1 models and 0.02 and 0.04 mm for the final Wing 2 model. The cross-sectional geometries and the shell thicknesses were also measured by an X-ray computer tomography scanner (TOSCANNER-32300μFPD) to evaluate the internal geometries, as shown in Figs. 15 and 16. The widths of the solid front section from the leading edge of the final wing models were 4.76 mm for the Wing 1 model on average and 5.67 mm for the Wing 2 model, which resulted in differences of 0.24 and 0.17 mm from the CAD models. The differences in average shell thicknesses on the upper and lower surfaces were 0.04 mm for the Wing 1 model and 0.08 mm for the Wing 2 model.

The surface roughness of each wing model was evaluated to ensure that the fabricated wing had enough smoothness for wind tunnel testing. Fig. 17 describes the measured locations of the surface roughness for the wing models. The surface roughness values were measured at the outer-, mid-, and inner-span on the upper and lower surfaces. The lines of 2.0 mm (Ra) in the chordwise direction were measured for the roughness evaluations. Measurements for each direction and point were performed five times. The averaged results are summarized in Table 4. The average roughnesses of the wing models on the line profiles (Ra) in the chordwise direction were mostly <1.0 μm for the upper and lower surfaces. The average roughness value around the mid-span of the Wing 2 model was higher than in the other regions. The higher roughness occurred due to a slight chatter during the machining process, which could be improved by fully optimizing the machining process parameters. However, the maximum surface roughness was still <3.5 μm. Therefore, we concluded that the surfaces were smooth enough for a wind tunnel test.

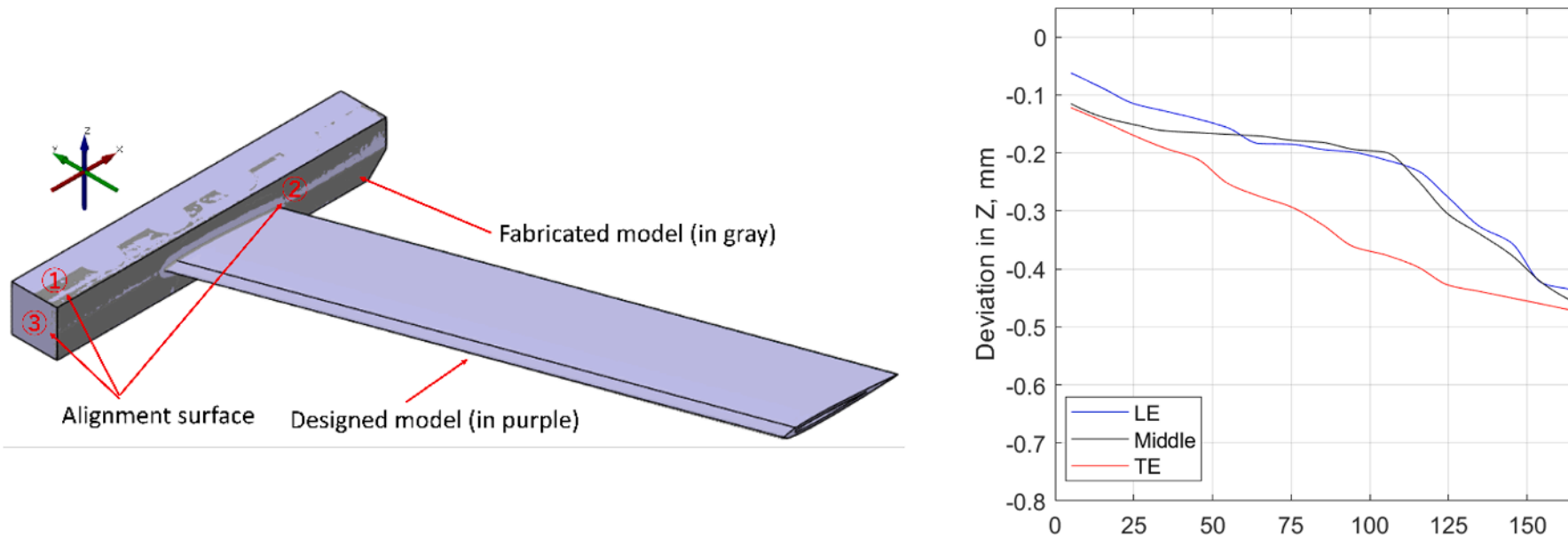

(a) The designed and fabricated wing models

(b) The surface deviation in the z-direction

**Fig. 10.** The comparisons of the final Wing 1a models and the surface deviation in the z-direction along the span at three chordwise locations.

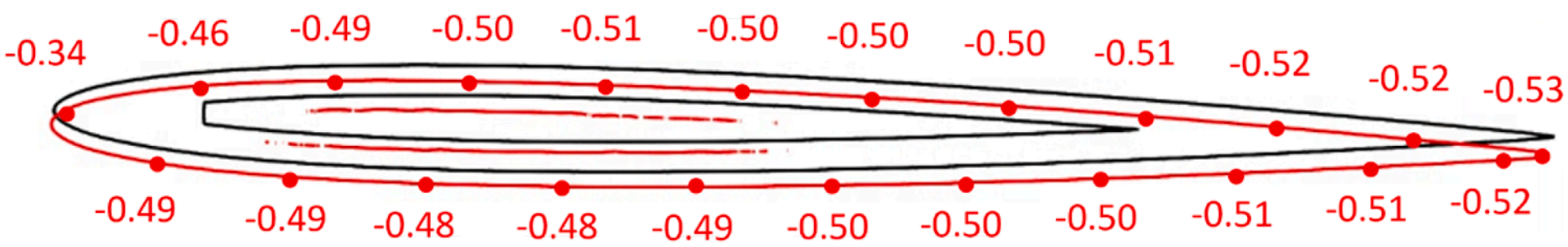

**Fig. 11.** The deviation of the cross-sectional geometry in the normal direction for the final Wing 1a model.

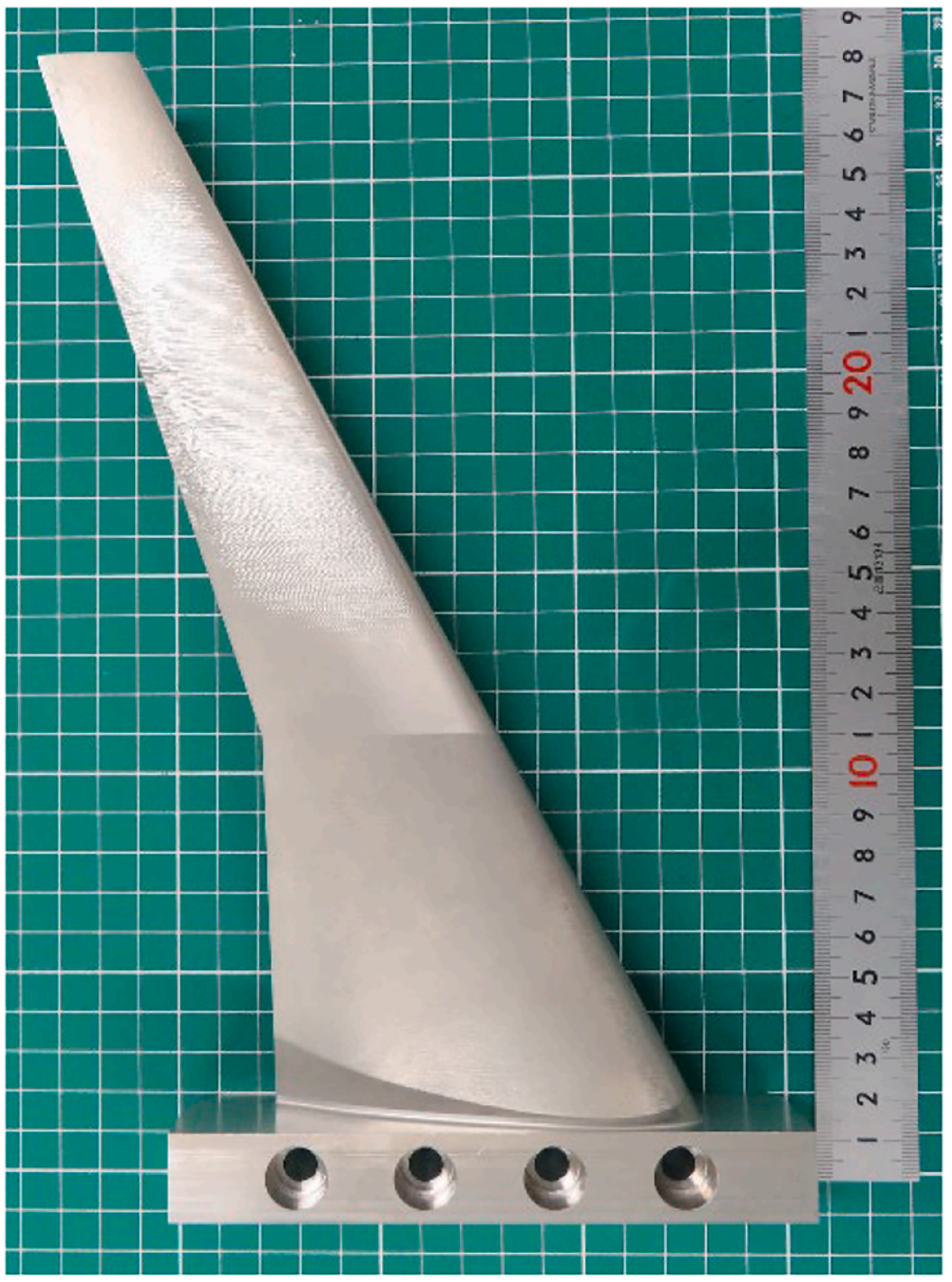

**Fig. 12.** A picture of the final model for the Wing 2 model.

## 4. Structural evaluation of fabricated wing models

This chapter describes structural evaluations of the wing models fabricated by the present combined processes. The static and dynamic characteristics of both wing models were evaluated with static load and vibration tests. Finite element (FE) simulations by MSC.Nastran were also performed to compare the numerical solutions with the experimental results obtained from the fabricated models.

### 4.1. Static characteristics

Static and modal simulations based on FE models were performed to evaluate the structural characteristics, such as stiffness and mass distributions, of the additively manufactured wing models with the present post-process technique. Fig. 18 describes the developed FE models of Wings 1 and 2. The wing roots were fixed for each model. The FE models of Wings 1 and 2 were divided into 3200 and 40,000 triangular shell elements, respectively, to accurately capture the characteristics of the wing models. Most individual shell elements on the wing surfaces had thicknesses of 1.0 mm for Wing 1 and 1.5 mm for Wing 2. Thicknesses of half the distance between the upper and lower surfaces were applied to the shell elements in the solid region and near the trailing edge. The wing was assumed to be an equivalent isotropic structure based on the properties along the layer direction. Our previous studies showed that this assumption was valid for predicting fundamental structural and aeroelastic characteristics of wing models [31,32,34].

Static load experiments with the fabricated wing models were performed to investigate the stiffness characteristics of the wings. The bases of the wing models were clamped to establish a cantilevered boundary condition. Tip loadings were applied at mid-chord: 0.9 N and 5.0 N for Wing 1, and 5.0 N and 9.9 N for Wing 2. These loadings were applied by installing calibrated weights. The vertical displacements of the wings were measured using laser displacement sensors (Keyence Corp.) at the locations shown in Fig. 19. The resolution of these sensors was 2 μm. Each measurement was obtained five times.

Comparisons of vertical deflections along the span for Wing 1 between the experimental measurements and numerical solutions are presented in Fig. 20 for different tip loads. In Fig. 20, the experimental data points represent the average vertical displacement of Wing 1a and

**Table 3**
External geometries of final wing models.

| Property | Wing 1 | | Wing 2 | |
|---|---|---|---|---|
| | Design | Fabricated | Design | Fabricated |
| Semi-span, mm | 200.00 | 200.04 (0.02 %) | 249.00 | 249.08 (0.03 %) |
| Root chord, mm | 50.00 | 50.06 (0.12 %) | 100.58 | 100.60 (0.02 %) |
| Tip chord, mm | 50.00 | 50.06 (0.12 %) | 23.14 | 23.18 (0.17 %) |

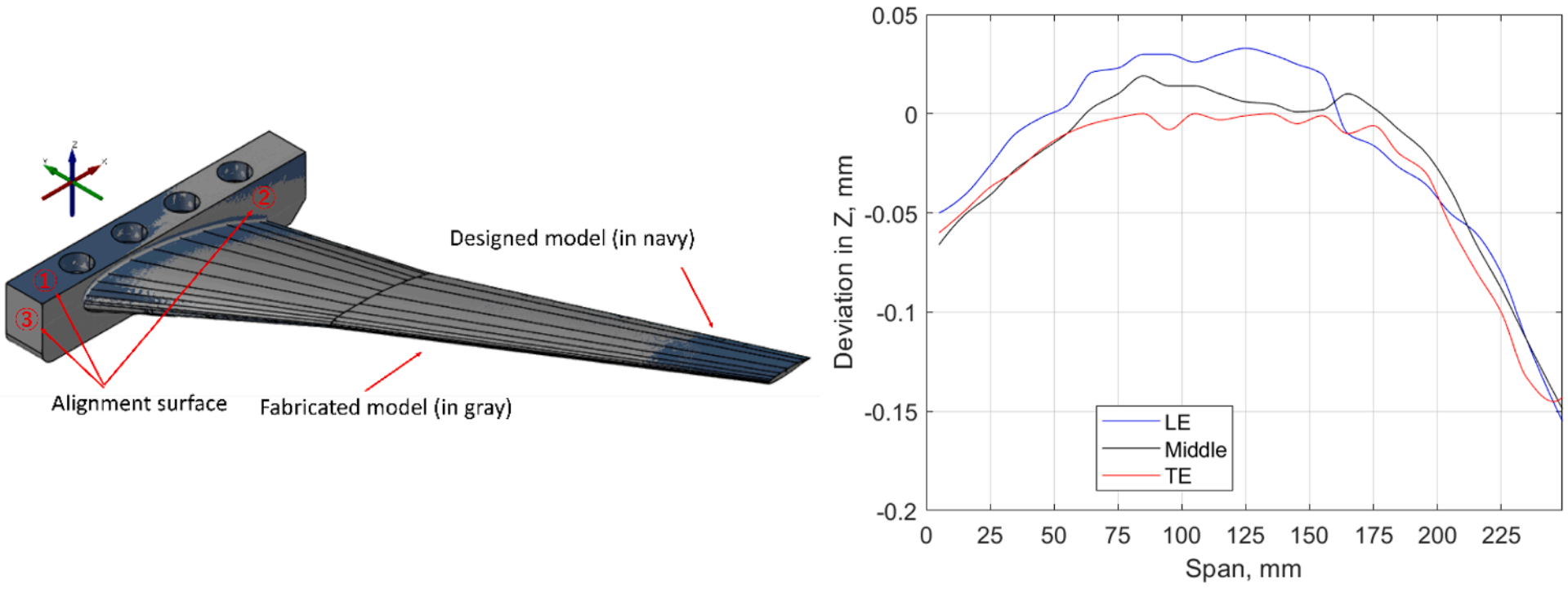

(a) The designed and fabricated wing models (b) The surface deviation in the z-direction

**Fig. 13.** The comparisons of the final Wing 2 models and the surface deviation in the z-direction along the span at three chordwise locations.

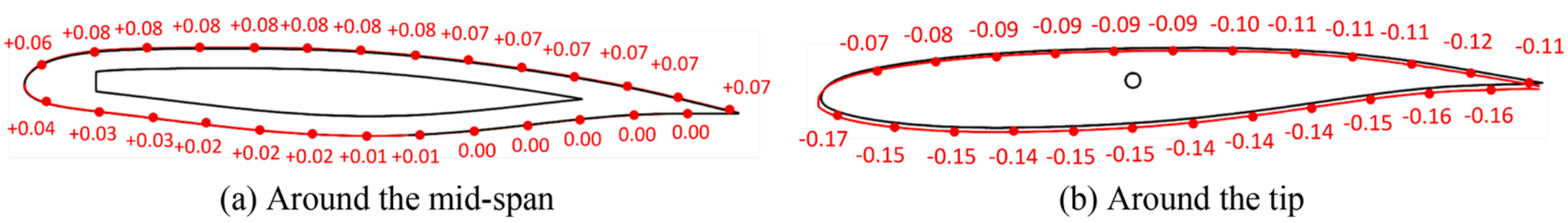

(a) Around the mid-span (b) Around the tip

**Fig. 14.** The deviation of the cross-sectional geometry in the normal direction for the final Wing 2 model.

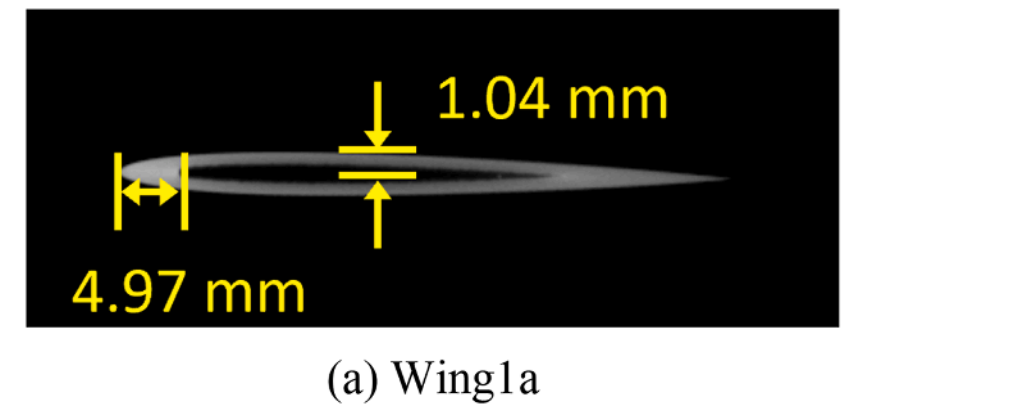

(a) Wing1a

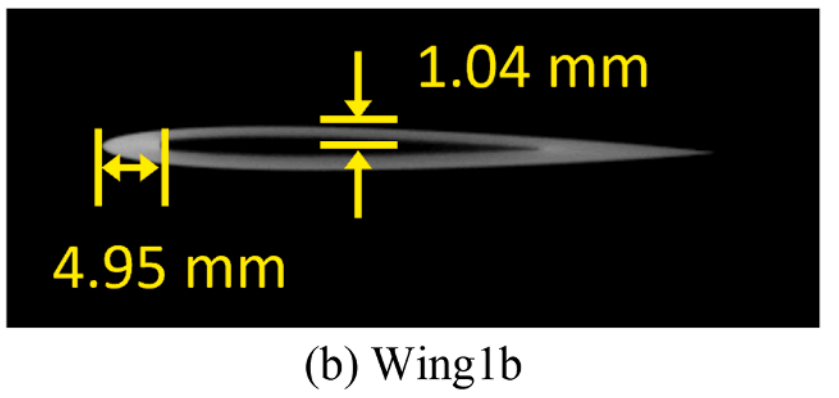

(b) Wing1b

**Fig. 15.** The cross-sections of the Wing 1 models obtained by the CT scan.

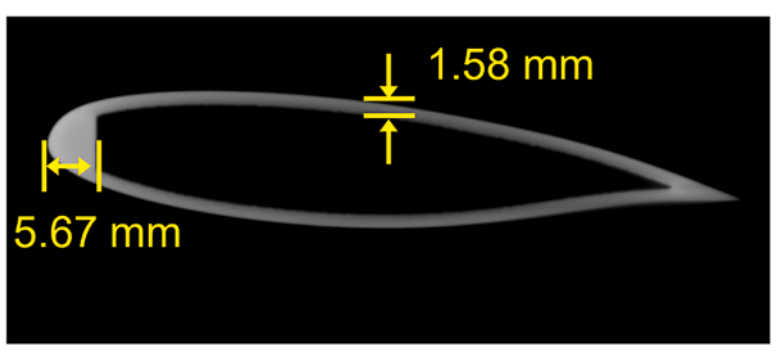

**Fig. 16.** The cross-section of the Wing 2 models obtained by the CT scan.

**Table 4**
Surface roughness measurements of the wing models on the upper and lower surfaces.

| Location ID. | Ra (Wing 1), μm | Ra (Wing 2), μm |
|---|---|---|
| 1 | 0.91 | 0.84 |
| 2 | 0.95 | 0.97 |
| 3 | 0.56 | 1.14 |
| 4 | 0.96 | 1.00 |
| 5 | 0.82 | 3.19 |
| 6 | 0.55 | 2.15 |

Wing 1b at each spanwise location. The error bars correspond to the standard deviation calculated from the measurements of these two identically designed models, thereby quantifying the reproducibility of the fabrication process. The fabricated Wing 1 model demonstrated good agreement with the numerical solutions for bending stiffness with average errors of 3.44 % and 3.78 % under 0.9-N and 5.0-N loading, respectively. The maximum standard deviation of the measured displacements for Wing 1 across all spanwise locations was 0.005 mm under 0.9-N loading and 0.003 mm under 5.0-N loading, taking into account the 2 μm resolution of the laser displacement sensors. This further confirms the consistent output across models.

Tables 5 and 6 summarize the comparisons of vertical deflections from the measurements and simulations for the more complex Wing 2 model. In these tables, the values in parentheses represent the standard deviation of the five repeated measurements at each location, indicating the consistency of the experimental data. Wing 2 exhibited a slightly larger difference from the numerical solutions compared to Wing 1. However, with average errors of 4.09 % and 4.19 % under 5.0- and 9.9-N loading, we concluded that the Wing 2 model was also constructed with sufficient accuracy to represent its designed stiffness characteristics.

### *4.2. Modal characteristics*

Linear modal simulations were then conducted to investigate the dynamic characteristics of wing models. Table 7 shows the simulated natural frequencies of the lower modes for the Wing 1 model. The lower out-of-plane (OOP) and torsional mode shapes of the Wing 1 model are shown in Fig. 21. The first OOP bending mode was the lowest mode with a natural frequency of 56.67 Hz, which was followed by the second OOP bending mode. After the first torsional mode with 419.22 Hz, there was the third OOP bending mode with a frequency of 730.79 Hz. The modal solutions were compared with the results of vibration tests in Table 7 to investigate the realizations of designed structural characteristics.

The GVT was performed to evaluate the dynamic characteristics of the wing models. The sampling frequency and the resolution of the GVT measurement were 5 kHz and 0.04 Hz. For Wing 1, two sets of GVT data obtained from the independently fabricated Wing 1a and 1b models (GVT (a) and GVT (b), respectively) are presented in Table 7. The modal characteristics of the fabricated Wing 1 model agreed with the solutions of the designed Wing 1 model with an average error of 2.16 %. The standard deviation between these two models for corresponding modes was up to approximately 1.4 Hz. This magnitude of deviation is clearly resolved by the GVT measurement, confirming that the observed variability reflects actual differences between the independently manufactured models. Despite this measurable variability, the consistency remains excellent, particularly considering the inherent complexity of aeroelastic model fabrication and the successful reproduction of consistent structural properties.

Table 8 lists the natural frequencies of the lower modes for the Wing 2 model obtained by the simulation and experiment. The lower OOP and torsional mode shapes of the Wing 2 model are shown in Fig. 22. Most frequencies obtained by the vibration test agreed with the numerical solutions, although the differences were slightly larger than those of Wing 1 due to the more sophisticated geometry. These differences for Wing 2 can be attributed to its higher geometrical complexities, such as the double-taper ratio and supercritical airfoil profile, which present greater challenges in precise manufacturing and replication of designed stiffness and mass distributions. Therefore, it was confirmed that the fabricated wing models based on the present manufacturing methodology accurately realized the modal structural characteristics of the designed models.

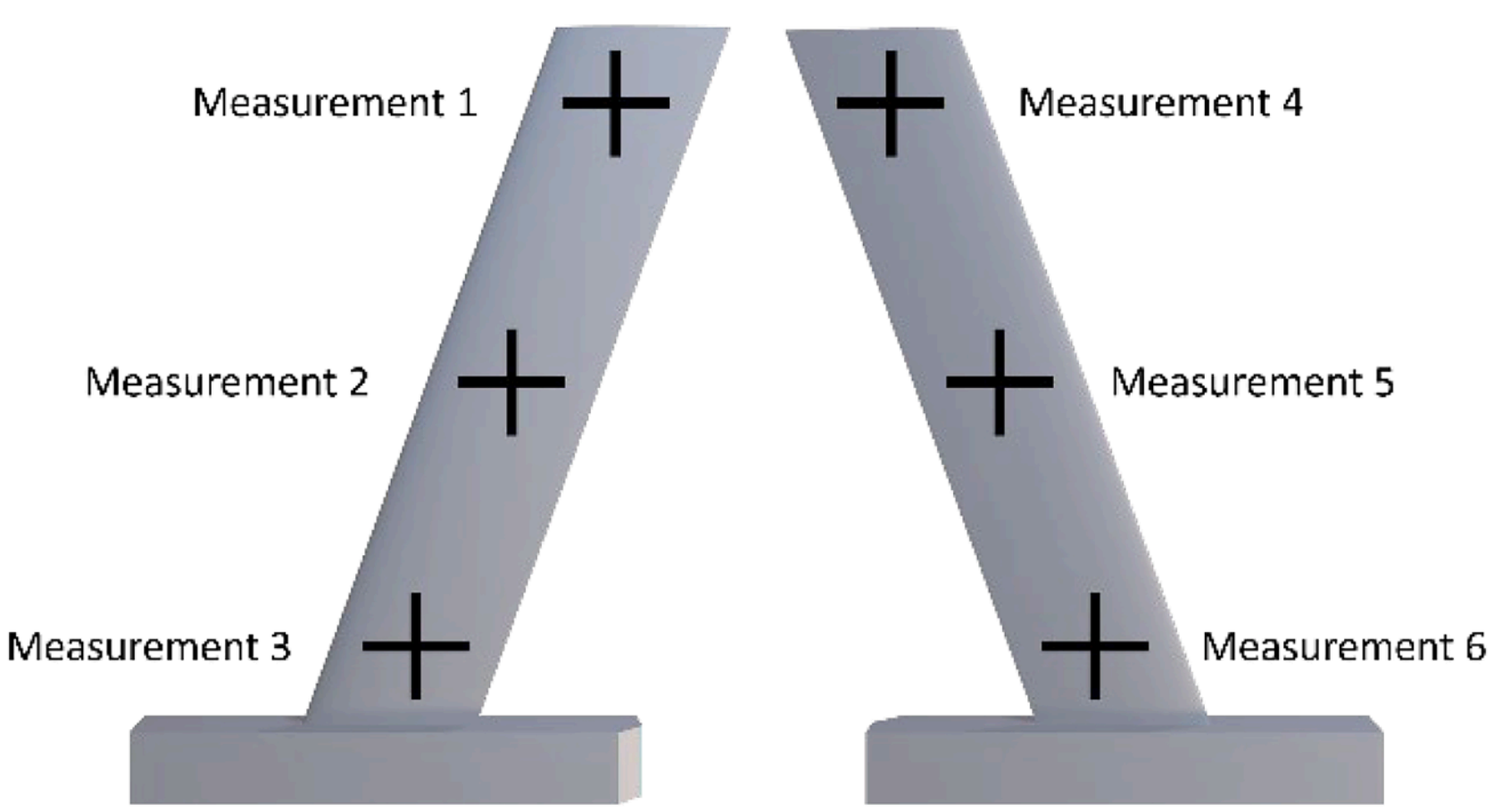

**Fig. 17.** Measured locations of surface roughness on wing models.

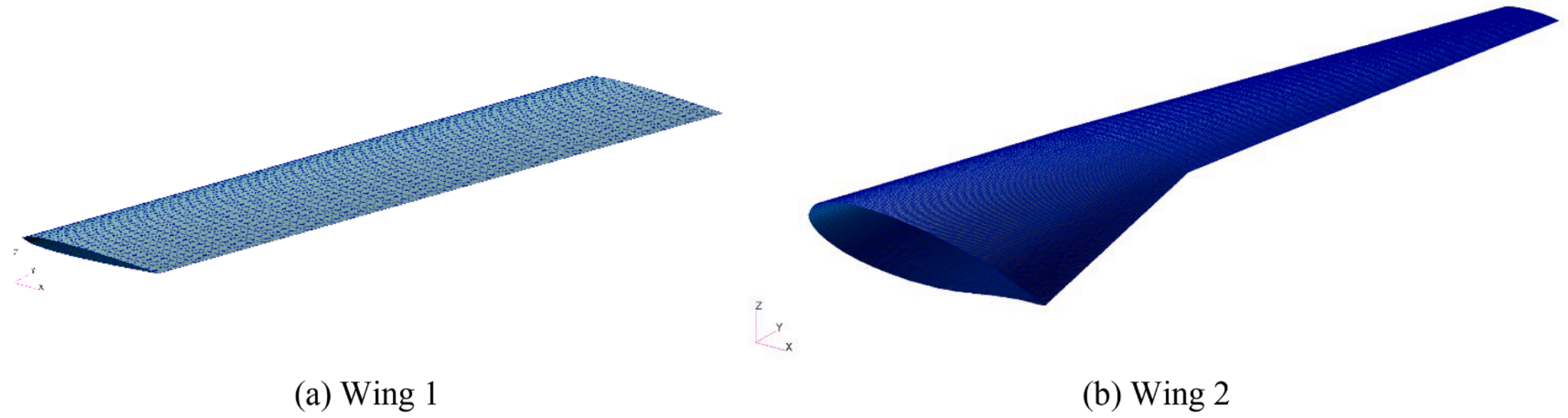

**Fig. 18.** Finite element models of Wings 1 and 2.

## 5. Transonic wind tunnel testing

Transonic flutter wind tunnel tests at the Transonic Flutter Wind Tunnel in Japan Aerospace Exploration Agency (JAXA) were performed to evaluate the aeroelastic characteristics of the fabricated Wing 1 models. The blow-down wind tunnel had a 0.6 $m$ × 0.6 m closed test section. Fig. 23 shows the wind tunnel setup. The total temperature in the wind tunnel was 19.1 °C. Two laser displacement sensors (Keyence Corp.) were used to measure the vertical deflections of the wing models on the upper surface. The first measurement location was 75.00 and 12.00 mm from the wing root and the leading edge, respectively, while the second location was 156.50 mm and 10.50 mm from the root and the trailing edge, respectively. The measurements were used to evaluate the modal characteristics in the GVT. Bending and torsional strains were also measured with the strain gauge rosettes attached near the wing root on the upper and lower surfaces. Vibratory aeroelastic responses were also measured using a high-speed camera (Redlake Corp.).

The flutter boundaries of the wing models obtained by the wind tunnel tests are given in Fig. 24. The flutter boundaries were determined based on observations of the flutter onset and the post-flutter phenomena. Wings 1a and 1b showed great agreement in the nonlinear aeroelasticity in the transonic regime (transonic dip). Figs. 25 and 26 show the aerodynamic conditions and displacement measurements for the runs where the aeroelastic instabilities were observed. Similar transient aeroelastic responses were observed in both wing models. The fast Fourier transform (FFT) analysis of the displacement measurements was also performed. The flutter frequencies for both Wing 1a and 1b were 157.0 Hz and 158.0 Hz, respectively, and they agreed very well. This close agreement in flutter frequencies from independently manufactured models strongly assets to the high reproducibility of the combined AM+SM fabrication methodology for aeroelastic characteristics. The difference of only 1.0 Hz is well within the typical experimental uncertainty for such complex aeroelastic phenomena. The high-speed camera observation also confirmed that the dominant vibration modes of the LCOs were the first OOP bending for both wing models, as shown in Fig. 27. These results collectively demonstrate that the combined manufacturing technique can fabricate flexible wind tunnel models with excellent reproducibility, enabling reliable and repeatable aeroelastic investigations.

## 6. Summary of model characteristics

To provide a comprehensive overview of the fabricated wing models' performance and to facilitate a direct comparison across different designs and evaluation metrics, this section summarizes the key geometrical, structural, and aeroelastic characteristics. The combined manufacturing approach aims to achieve high precision and reproducibility, and Table 9 quantifies these aspects.

## 7. Discussion

The preceding sections have demonstrated the efficacy of the combined AM and SM approach for fabricating high-quality, flexible wing models for transonic wind tunnel testing. This section further discusses the implications of our findings, addresses specific points raised by the reviewers, and outlines future research directions.

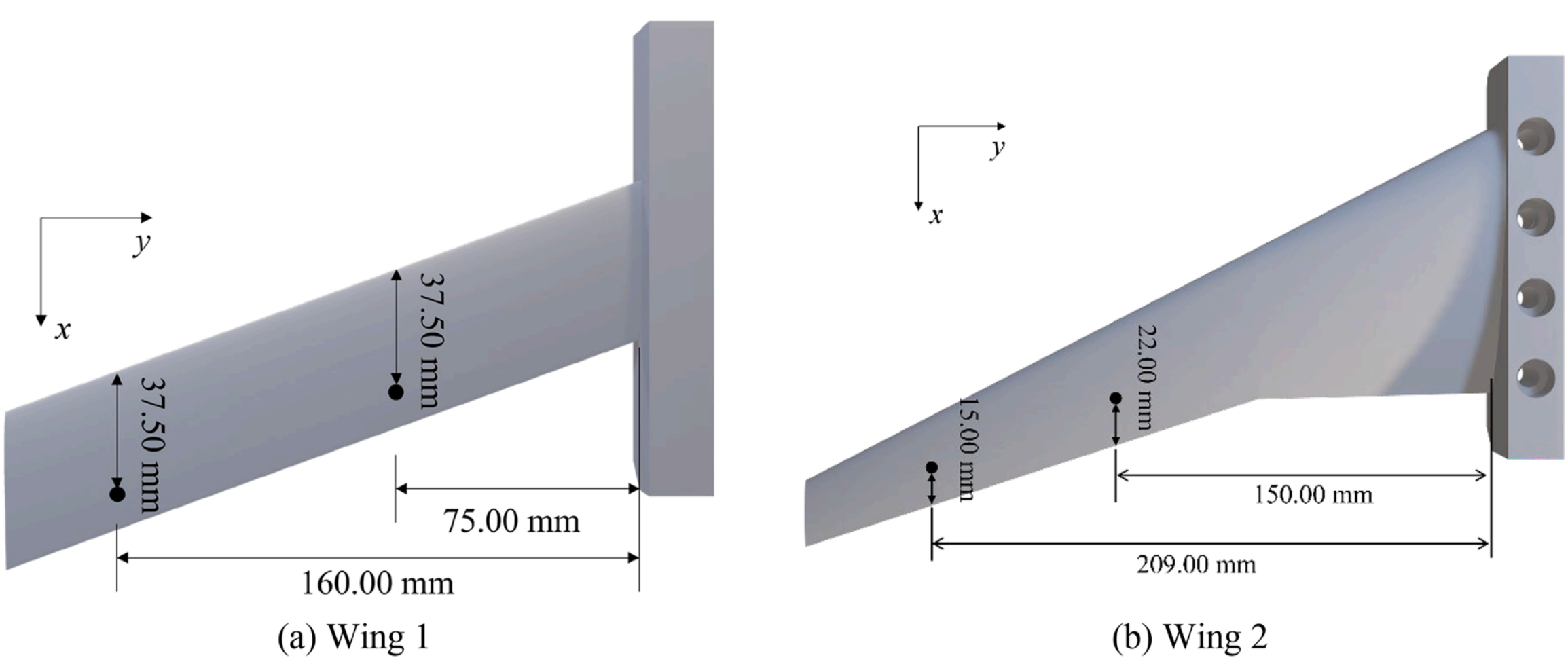

**Fig. 19.** Measurement locations for Wings 1 and 2 models.

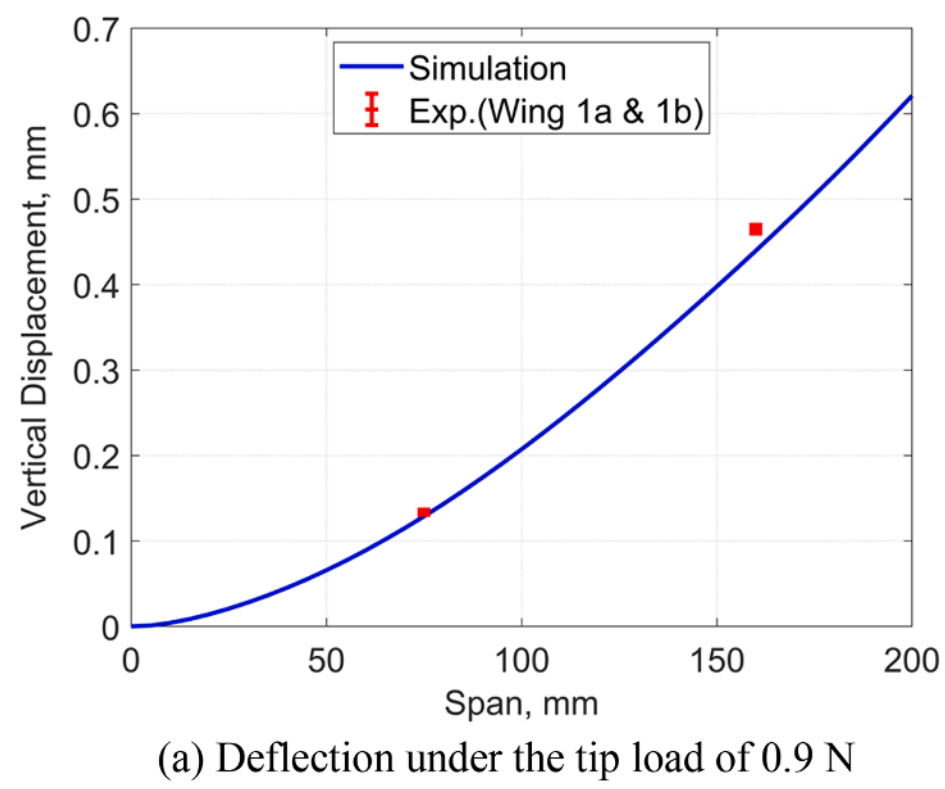

(a) Deflection under the tip load of 0.9 N

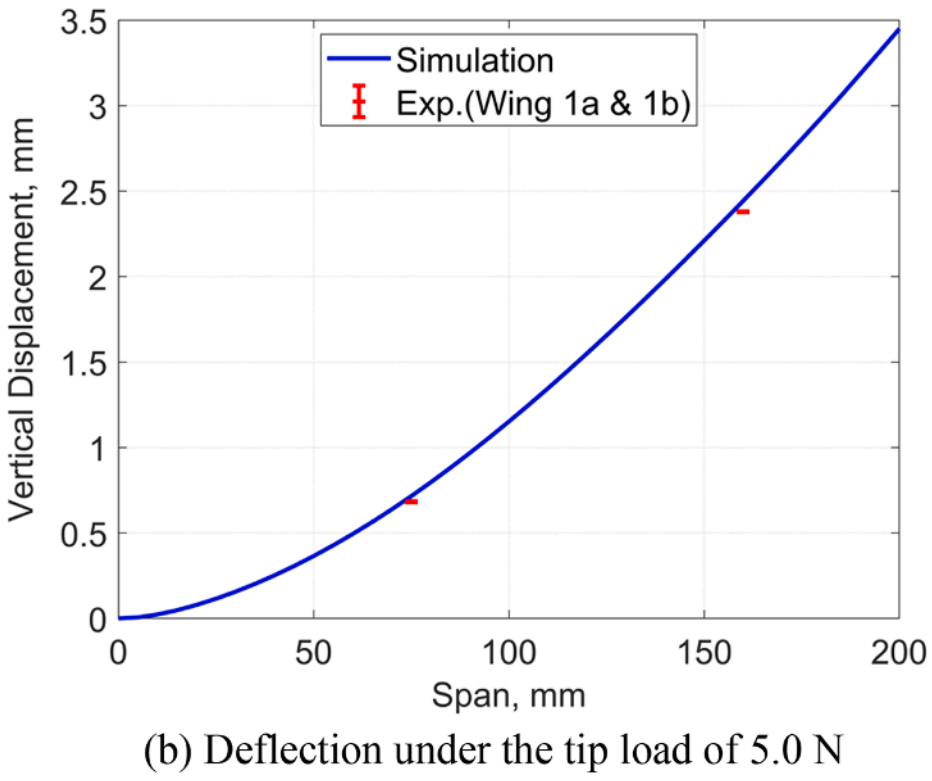

(b) Deflection under the tip load of 5.0 N

**Fig. 20.** Vertical displacements along the span for the Wing 1 models from the experiment and simulation under different tip loads.

**Table 5**
Vertical displacements for wing 2 under 5.0 N.

| Wing | FE model | | Fabricated | |
|---|---|---|---|---|
| Spanwise pos., mm | 150 | 209 | 150 | 209 |
| Chordwise pos., mm | 22 | 15 | 22 | 15 |
| Deflection, mm | 0.1435 | 0.4614 | 0.1445 (0.001) | 0.4269 (0.001) |

### 7.1. Novelty and benchmarking of the combined AM + SM approach

While AM-based fabrication of wind tunnel models is not new, the systematic integration of AM with precision SM, as presented in this study, significantly enhances the current state-of-the-art. Previous efforts, including our own, often relied on skill-dependent post-processing methods such as MP to achieve desired surface finishes. While AM-only methods offer rapid prototyping, they typically result in insufficient surface roughness and dimensional accuracy for high-fidelity aerodynamic studies. Mechanical polishing, although improving surface finish, introduces variability and skill dependency, leading to inconsistencies in the final model's quality and, consequently, in experimental data reproducibility.

The novelty of the present AM + SM approach lies in its ability to systematically provide stable quality and enhanced reproducibility across multiple fabricated models. The CNC machining step ensures precise control over the final geometry and surface roughness, mitigating the human error and variability inherent in manual post-processing. This is evidenced by the consistent flutter boundaries observed between Wing 1a and Wing 1b (Fig. 24), with flutter frequencies of 157.0 Hz and 158.0 Hz, respectively. This remarkable agreement underscores the improved reproducibility of aeroelastic behavior across different models manufactured with this methodology. Furthermore, the average surface roughness achieved was <1.0 μm, and the average surface deviation was <0.3 mm, representing a notable improvement in geometrical precision compared to our previous methods.

To further contextualize the benefits of our approach, Table 10 provides a comparative analysis of various wing model fabrication methods in terms of cost, lead time, geometrical accuracy, surface roughness, and quality reproducibility. As shown in Table 10, while conventional methods can achieve high accuracy, they often suffer from high costs and long lead times, especially for complex internal geometries inherent to aeroelastic scaling. AM-only fabrications offer speed but sacrifice surface quality. Our AM + SM method strikes a balance, providing superior accuracy and surface finish compared to prior AM approaches while maintaining the benefits of AM's design freedom for internal structures at a competitive cost and lead time. This systematic approach contributes significantly to the efficient and reliable production of wind tunnel models.

**Table 6**
Vertical displacements for wing 2 under 9.9 N.

| Wing | FE model | | Fabricated | |
|---|---|---|---|---|
| Spanwise pos., mm | 150 | 209 | 150 | 209 |
| Chordwise pos., mm | 22 | 15 | 22 | 15 |
| Deflection, mm | 0.2842 | 0.9136 | 0.2920 (0.002) | 0.8621 (0.002) |

**Table 7**
Lower natural frequencies of Wing 1 with respect to bending and torsional modes.

| Mode ID | Mode | Simulation, Hz | GVT (a), Hz | GVT (b), Hz |
|---|---|---|---|---|
| 1 | 1st out-of-plane bending | 56.67 | 58.0 | 60.0 |
| 2 | 2nd out-of-plane bending | 304.39 | 293.0 | 305.5 |
| 3 | 1st torsion | 419.22 | 408.0 | 424.0 |
| 4 | 3rd out-of-plane bending | 730.79 | 732.0 | 728.0 |
| 5 | 1st edgewise bending | 769.45 | – | – |

### 7.2. Effects of residual stress and potential variability

The geometrical evaluation of the final models indicated slight warping, particularly near the wing tip (Fig. 11). This warping can be partially attributed to the residual stresses inherently present in additively manufactured structures. While the SM process effectively refines the external geometry and surface finish, it can also relieve some of these internal stresses, potentially inducing further deformation in the final part. Although the overall geometrical precision was significantly improved by the present approach, the influence of this subtle warping on modal accuracy and aeroelastic behavior merits further consideration. Specifically, while minor static OOP bending deformations may be tolerable, residual stresses could potentially induce torsional deformations, thereby altering the effective angle of attack across the wing span, which is critical for aeroelastic characteristics.

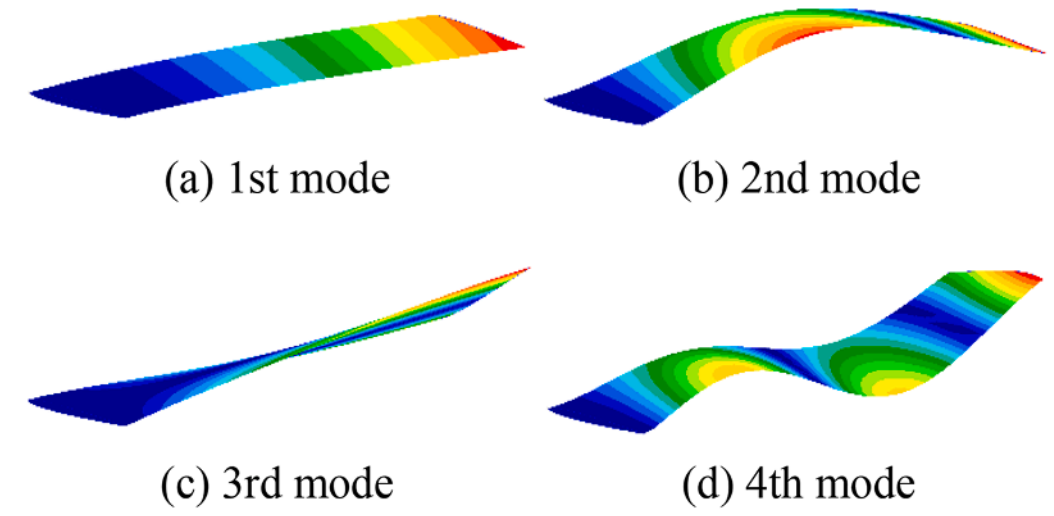

(a) 1st mode (b) 2nd mode

(c) 3rd mode (d) 4th mode

**Fig. 21.** The lower out-of-plane and torsional mode shapes for the Wing 1 model.

The presence of internal residual stresses can also influence the apparent stiffness of a structure. For instance, compressive residual stresses can enhance resistance to crack initiation and propagation, effectively increasing the load-bearing capacity and, in some contexts, the perceived stiffness against external tensile loads. If this influence leads to an increase in the overall structural rigidity, it could potentially raise the natural frequencies of vibration modes, which in turn might lead to an increase in the flutter speed. However, an increase in stiffness, while often raising flutter speed, does not guarantee it. Flutter instability frequently arises from the frequency coalescence of multiple vibration modes. Consequently, changes in stiffness that alter the frequency separation between interacting modes can either increase flutter speed (if the separation widens, delaying coalescence) or decrease it (if the separation narrows, accelerating coalescence). For Wing 1, the impact of this residual stress-induced warping on modal characteristics was minimal, as evidenced by the good agreement between experimental and numerical results (Table 7). For Wing 2, however, slight discrepancies in natural frequencies were observed (Table 8), which could be partially linked to its more complex geometry and potentially larger residual stress effects. The fact that Wing 1a and 1b showed excellent agreement in flutter boundaries (Fig. 24) suggests that the SM process, by ensuring a consistent final geometry, effectively mitigates the variability that residual stresses might otherwise introduce into the aeroelastic response. This indicates that while residual stress may cause some static deformation, its dynamic impact, including potential torsional effects and associated changes in local angle of attack, is substantially controlled by the precision of the subtractive manufacturing step.

The choice of AM build orientation is another critical manufacturing parameter influencing the material properties and, consequently, the aeroelastic behavior of the fabricated models. For the precise replication of aerodynamically critical airfoil shapes, building the wing section by stacking layers in the spanwise direction (as shown in Fig. 4) is often chosen for surface accuracy. However, AM processes typically exhibit material anisotropy, where properties (such as stiffness and strength) can differ significantly between the build direction and directions within the build plane. Stiffness and strength along the build direction can be lower compared to in-plane directions. The introduction of the SM process in our methodology allows for the precise reproduction of the airfoil geometry regardless of the initial as-built surface quality, potentially offering more flexibility in choosing build orientations that optimize structural properties. For example, fabricating the model in a different orientation might lead to increased stiffness in the spanwise direction, which would typically contribute to an increase in flutter speed. However, similar to the discussion on residual stress, the effect on flutter speed is not straightforward, as it also depends on how the increased spanwise stiffness alters the frequency relationship between various coupled modes, such as bending and torsion, thereby affecting their frequency coalescence behavior.

Furthermore, the powder composition used in AM, AlSi10Mg in this study, plays a crucial role in the final material properties and thus the structural and aeroelastic characteristics. Variations in powder quality, particle size distribution, or impurities can lead to microstructural differences, affecting mechanical properties such as elasticity, strength, and ductility, which in turn influence the model's stiffness and damping. While the consistency of commercial AlSi10Mg powder typically ensures reasonable uniformity, subtle variations could contribute to inter-model discrepancies in dynamic response. Future work will involve a more detailed characterization of the as-built material properties, considering the influence of both process parameters and powder batch variability, to further refine the predictive capabilities for aeroelastic models.

**Table 8**
Lower natural frequencies of Wing 2 with respect to bending and torsional modes.

| Mode ID | Mode | Simulation, Hz | GVT, Hz |
|---|---|---|---|
| 1 | 1st out-of-plane bending | 139.49 | 156.0 |
| 2 | 2nd out-of-plane bending | 508.47 | 580.0 |
| 3 | 1st torsion + 2nd out-of-plane bending | 912.17 | 940.0 |
| 4 | 1st torsion + 2nd out-of-plane bending | 1036.9 | 1064.0 |
| 5 | 1st edgewise bending | 1151.8 | 1160.0 |

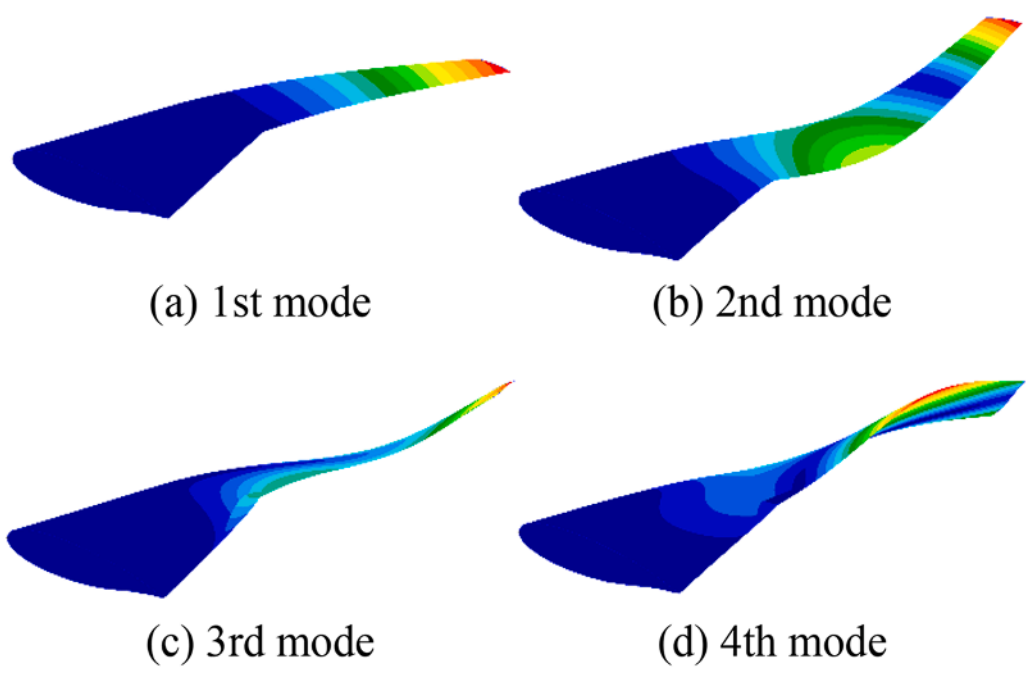

**Fig. 22.** The lower out-of-plane and torsional mode shapes for the Wing 2 model.

Future research will focus on optimizing AM process parameters, such as build orientation and post-build heat treatments, in conjunction with the machining process, to further minimize residual stress and its impact on final model geometry and structural properties. This includes investigating potential variability due to powder composition and its subtle effects on flutter boundaries, although the high reproducibility observed in this study suggests these effects are currently within acceptable limits for practical applications.

### *7.3. Applicability of the methodology and future directions*

This study successfully demonstrated the applicability of the AM + SM methodology for two distinct wing designs. While Wing 1, a simpler rectangular swept wing, showed excellent agreement with FE predictions in all aspects, Wing 2, a more complex double-tapered swept wing with a supercritical airfoil, exhibited slightly larger deviations from FE predictions for certain modal frequencies (Table 8). Further detailed investigation of Wing 2′s cross-sectional geometry using X-ray CT scans (as partially presented in Fig. 16) is planned to better understand these discrepancies and refine our modeling approaches. The increased complexity of Wing 2, encompassing sweep, double-taper ratio, and cambered airfoil profiles, presented a greater challenge in precisely replicating the designed stiffness and mass distributions, even with the enhanced accuracy of the AM + SM process. This suggests that for highly intricate geometries, further optimization of both AM build parameters and SM strategies may be required to fully mitigate subtle manufacturing imperfections that accumulate across complex features.

For Wing 1 and Wing 2, the average errors in shell thickness were 4.0 % and 5.3 %, respectively, at the CT scan measurement spanwise locations. The design of Wing 2 simulates a practical configuration, incorporating a double taper and a more complex supercritical airfoil compared to a typical NACA airfoil. This positions the Wing 2 design as approaching a worst-case scenario for conventional wing design in terms of geometric complexity. Despite this, the increase in manufacturing error compared to Wing 1 was limited to approximately 1.3 %, which strongly indicates the high applicability of the proposed method. For simpler tapered wings and airfoils, a higher probability of fabricating models within these geometrical error ranges is expected, thereby promising excellent reproducibility of their dynamic characteristics.

The present methodology holds significant promise for fabricating a broader range of aeroelastic test articles beyond simple wing models.

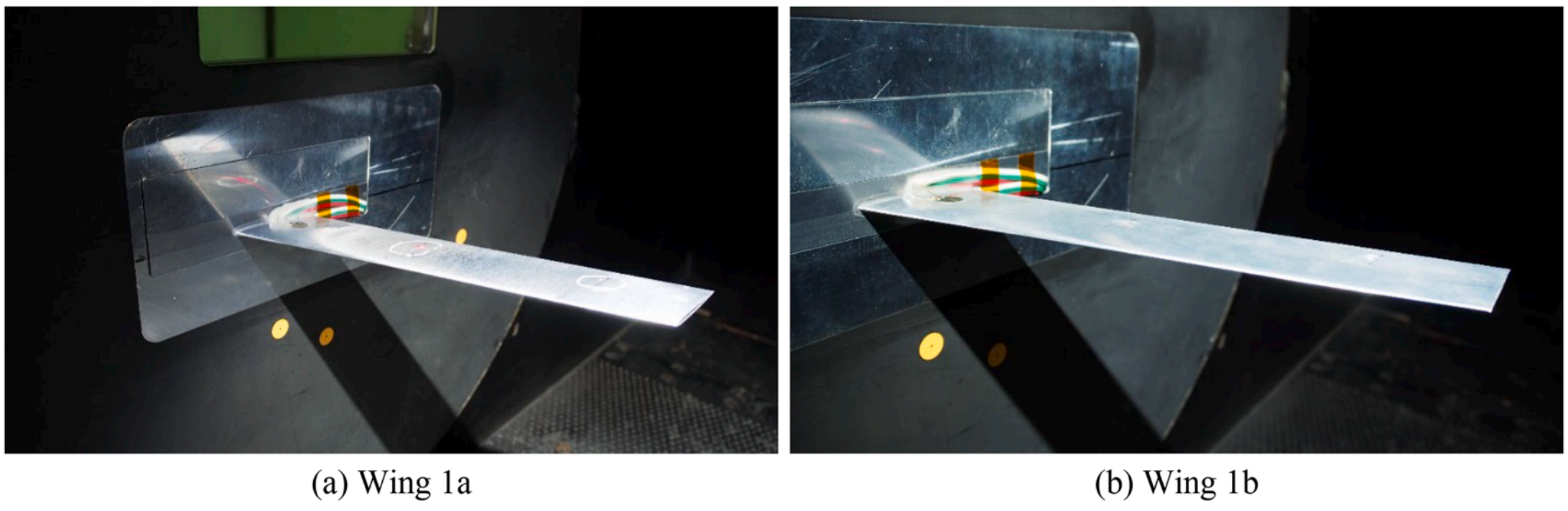
(a) Wing 1a (b) Wing 1b

**Fig. 23.** Wind tunnel test setup for two Wing 1 models.

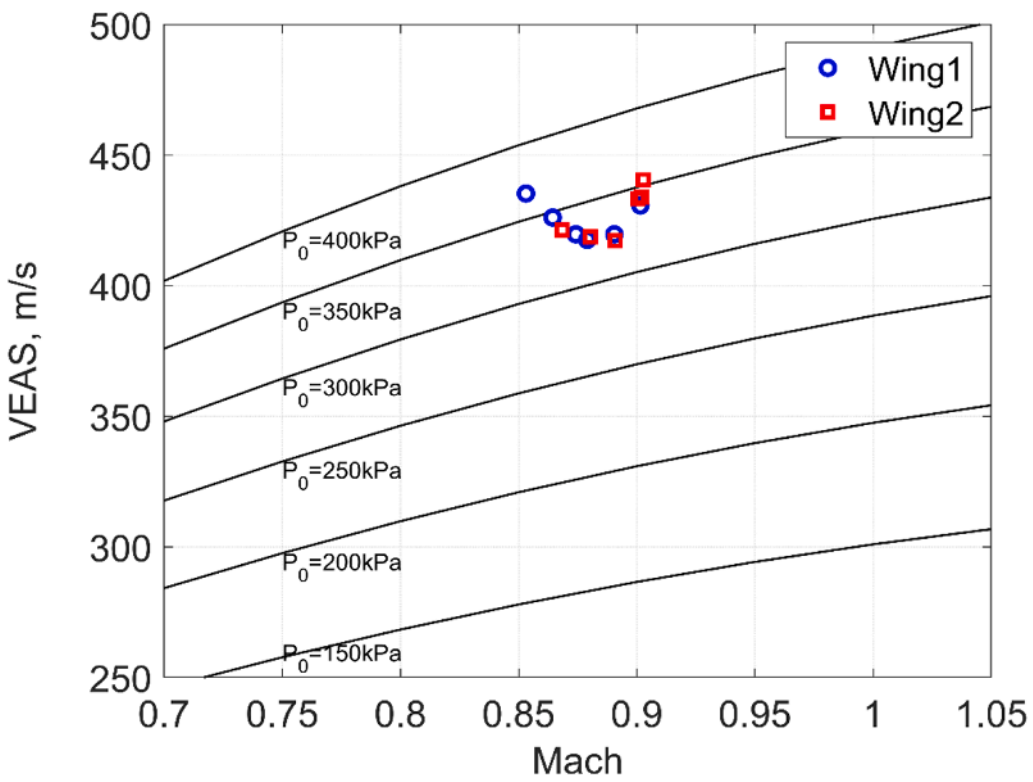

**Fig. 24.** Flutter boundary for the wing models.

This includes:

1. Control Surfaces: The precision offered by SM is crucial for control surfaces, where accurate hinge lines, tight tolerances, and faithful aerodynamic profiles are paramount for realistic aerodynamic and aeroelastic responses. The combined AM+SM approach enables the creation of lightweight internal structures via AM, followed by high-precision external shaping essential for these critical components. Specifically, this technique facilitates the design and fabrication of control surfaces with precisely defined hinge mechanisms and internal stiffeners, which can be challenging to achieve with conventional manufacturing. For instance, the AM process allows for the creation of complex internal geometries necessary to house control mechanisms such as servos and motors, achieving both weight reduction and the required stiffness and strength. The subsequent SM ensures that the external aerodynamic profile and critical interfaces, such as the hinge line, are accurately realized, minimizing aerodynamic penalties and ensuring realistic aeroelastic responses.
2. Integrated Sensor Models: The design freedom of AM allows for the seamless integration of internal cavities and conduits for various sensors (e.g., pressure taps, accelerometers, fiber optic strain gauges). The subsequent SM process can then ensure precise sensor housing, alignment, and external surface quality, enabling the creation of "smart" models for advanced experimental techniques. This capability reduces the need for external wiring and surface-mounted sensors, minimizing aerodynamic interference. As part of ongoing efforts, the integration of Fiber Optic Sensors (FOSs) into AM-fabricated wing models is being explored for enhanced structural monitoring. While current practices for FOS integration often involve embedding during the layer-by-layer AM process, which can limit sensor placement to the build plane, our approach, complemented by precision machining, offers the potential for more versatile sensor channel designs and improved integration quality. This will facilitate advanced shape identification based on strain measurements [35–38] for more accurate deformation monitoring in complex transonic flow environments.
3. Full-Span Half-Models: While our current study focused on cantilevered half-models, the methodology is scalable to larger or full-span half-models. Potential challenges include limitations of current AM printer build volumes and the machining envelope required for larger components. Maintaining structural integrity during machining for larger, more flexible parts also presents a challenge that would need to be addressed through careful design, support structures, and process planning. Despite these challenges, the ability

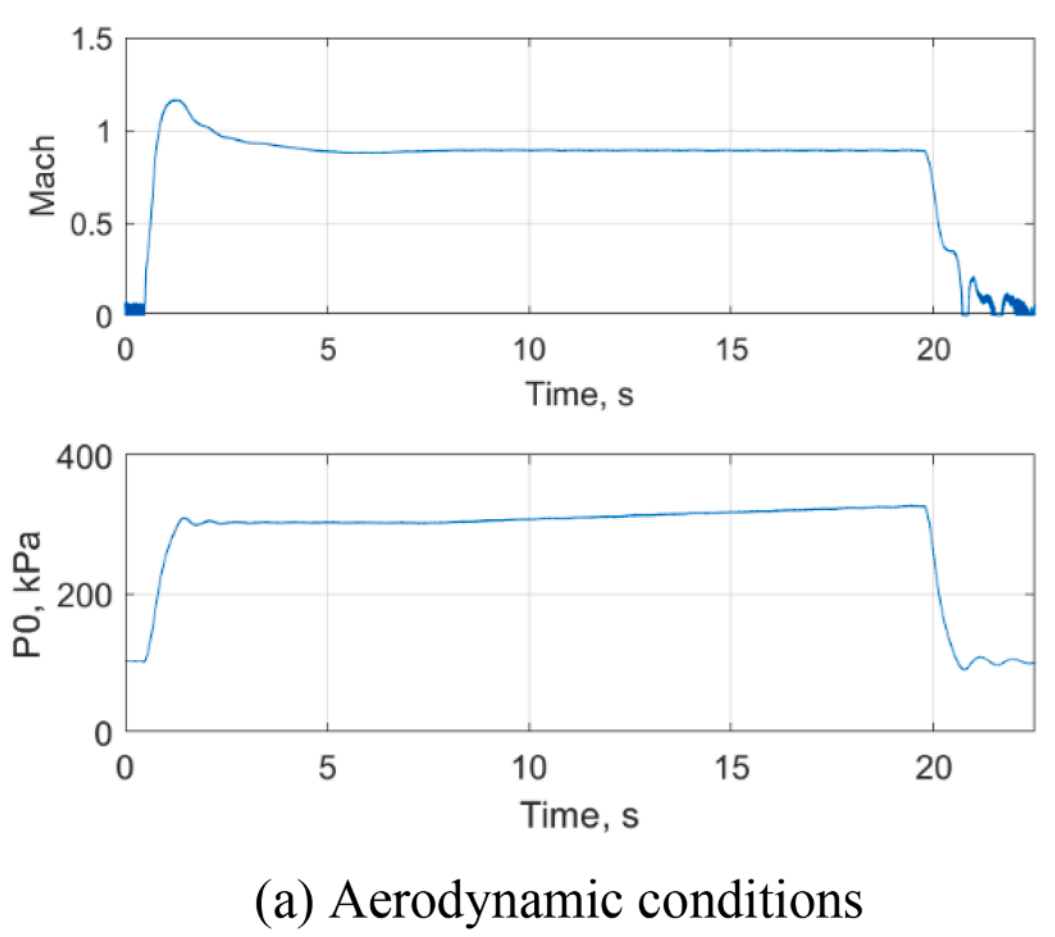

(a) Aerodynamic conditions

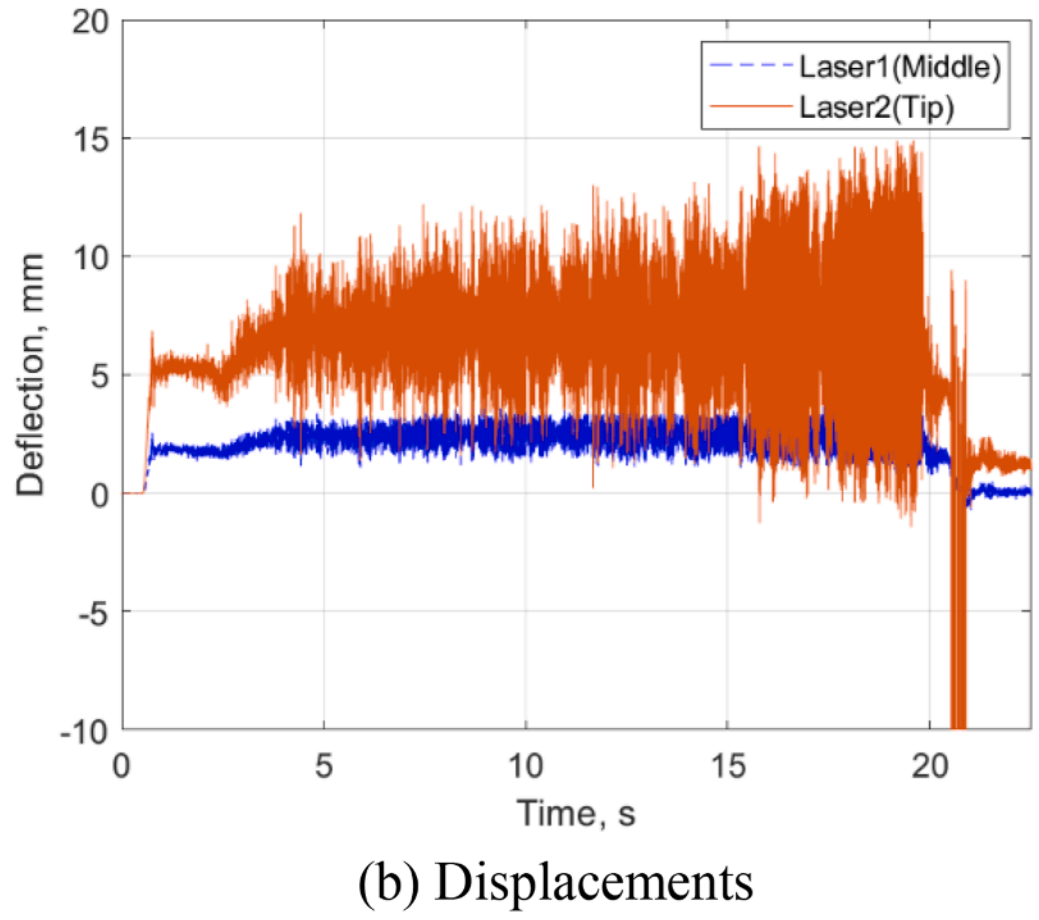

(b) Displacements

**Fig. 25.** Aerodynamic conditions and displacement measurements (right) for Wing 1a at $M = 0.89$ with a $P_0$ sweep.

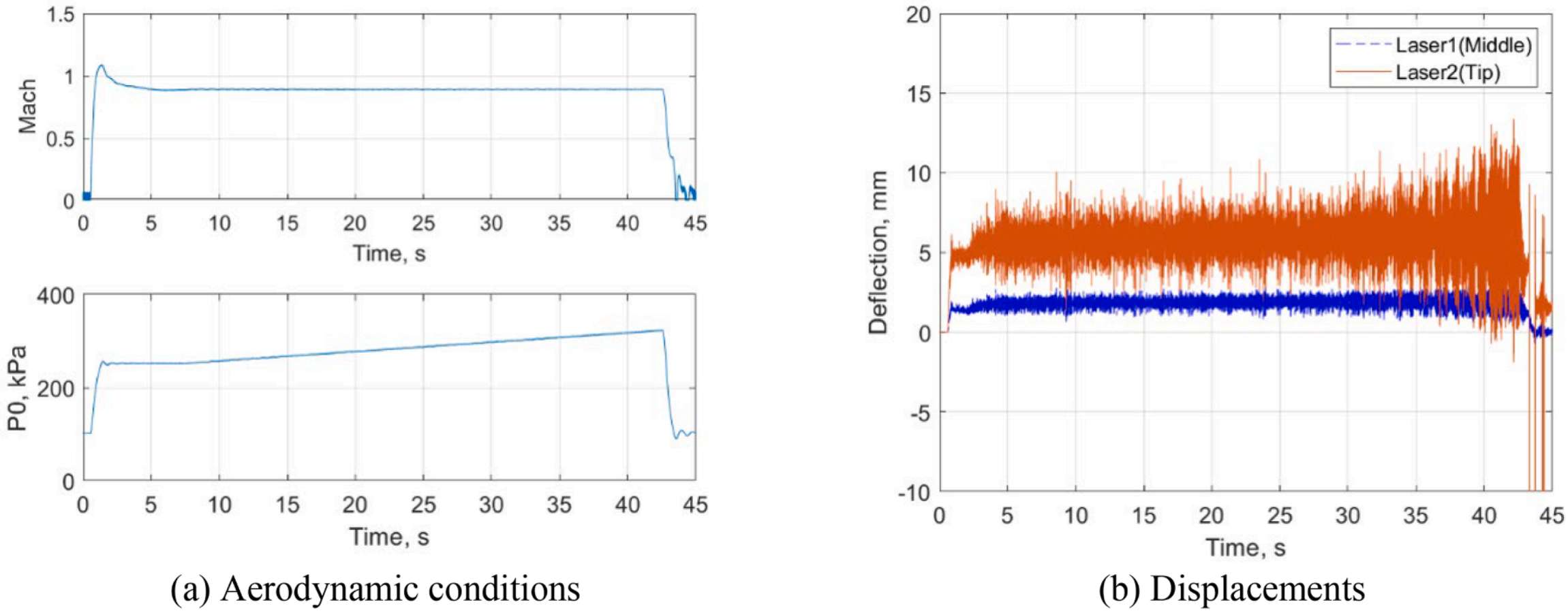

**Fig. 26.** Aerodynamic conditions and displacement measurements for Wing 1b at $M = 0.89$ with a $P_0$ sweep.

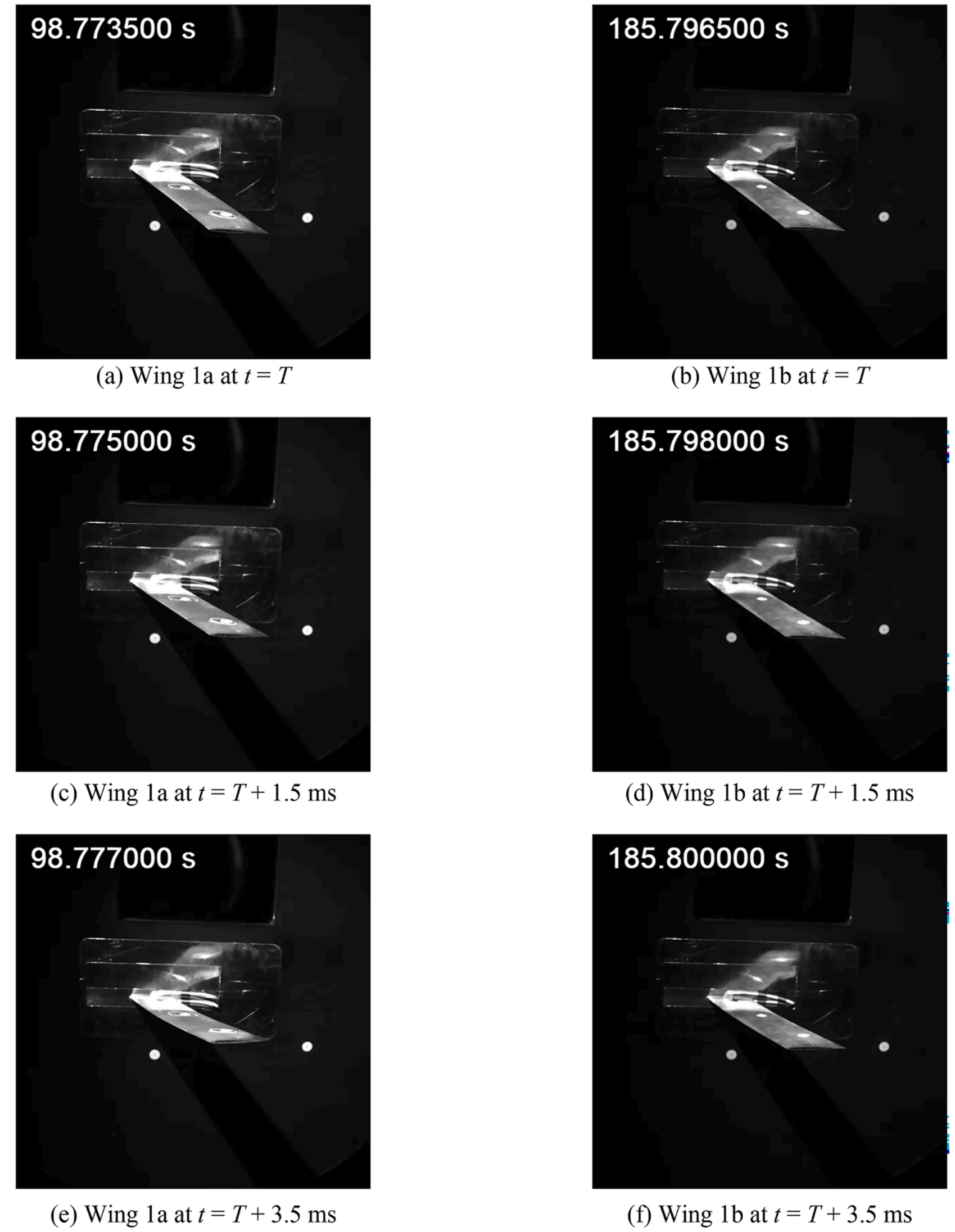

(a) Wing 1a at $t = T$ (b) Wing 1b at $t = T$
(c) Wing 1a at $t = T + 1.5$ ms (d) Wing 1b at $t = T + 1.5$ ms
(e) Wing 1a at $t = T + 3.5$ ms (f) Wing 1b at $t = T + 3.5$ ms

**Fig. 27.** Snapshots of the high-speed camera observations for Wing 1a and 1b during the aeroelastic instabilities.

to rapidly produce complex internal geometries (e.g., spars, ribs, and internal ducts) with AM, followed by precise external finishing, offers a compelling advantage over traditional fabrication methods for larger models.

Further applications could include complex lattice-structured wings, which leverage AM's unique capabilities for lightweight yet stiff designs. Recent work has extensively explored the rapid prototyping of aeroelastic test articles [26,27] and lattice structures for aerospace applications [24,25]. Our methodology complements these efforts by providing a robust framework for achieving the high precision often required for quantitative aeroelastic studies, building upon the foundational design freedoms offered by AM.

**Table 9**
Comparative analysis of key wing model characteristics and fabrication accuracies.

| Metric | Wing 1 (Averages of 1a and 1b) | Wing 2 | Reference [31] | Note |
|---|---|---|---|---|
| **Geometrical accuracy** | | | | |
| Average surface roughness (Ra), $\mu$m | < 1.0 | < 1.1 (max 3.5 due to chatter) | ~1.1 | Achieved by the SM process for aerodynamics |
| Average surface deviation (overall), mm | < 0.3 (max tip: ~0.5) | 0.03 (max tip: ~0.2) | ~0.16 (for a flutter wing model) or ~0.3 (for other designs) | Improved and stable precision by the present AM+SM method |
| Shell thickness error (average) | 4.0 % | 5.3 % | N/A | Increase for complex geometry, but still low |
| **Structural characteristics** | | | | |
| Static deflection error (average) | 3.44 % - 3.78 % | 4.09 % - 4.19 % | 2.38 % (for a flutter wing model) | Good agreement with FE solutions |
| Modal frequency error (average) | 2.16 % | Slightly larger (see Table 8) | < 4 % (for a flutter wing model) | Wing 2 shows a larger discrepancy for higher modes |
| **Aeroelastic characteristics** | | | | |
| Flutter frequency, Hz | 157.0 (1a) / 158.0 (1b) | Not tested | 187.0 (Only one model) | High reproducibility demonstrated for Wing 1 |

**Table 10**
Comparative analysis of wing model fabrication methods.

| Fabrication method | Cost | Lead time | Geometrical accuracy | Surface roughness | Quality reproducibility | Notes |
|---|---|---|---|---|---|---|
| CNC-only | Medium-High | Medium-Long | High | Very Low | High | Limited design freedom for internal structures, complex for thin-walled parts |
| Traditional composites layup | High | Long | Medium-High | Medium | Medium-High | Labor-intensive, difficult for complex internal geometries |
| AM-only (as-built) | Low-Medium | Short | Medium | High (rough) | Medium | Requires post-processing for aerodynamic surfaces |
| AM+MP | Medium | Medium | Medium-High | Low | Low (skill-dependent) | Highly dependent on manual skill, high variation rates |
| AM+SM (present) | Medium | Medium | High (improved) | Very Low (< 1.0 μm) | Very High | Automated, systematic, reduces human error and variability, enables repeatable test data |

### *7.4. Uncertainty quantification*

In this study, measurements, such as surface roughness and static deflections, were performed multiple times (e.g., five times), and average values were reported. For static deflections, error bars, representing the standard deviation of measurements from the two identically fabricated Wing 1 models (Wing 1a and Wing 1b) in Fig. 20, demonstrate the reproducibility of the manufacturing process. For frequency comparisons, the resolution of the GVT measurement was 0.04 Hz, and consistency between repeated GVTs (Wing 1a and 1b) demonstrated high reproducibility (Table 7).

While the reported mean values and reproducibility metrics provide confidence in the data, a more rigorous and comprehensive uncertainty quantification analysis, including all sources of experimental uncertainty (e.g., sensor calibration, environmental conditions, model fixturing), would further strengthen the results. This will be a focus for future experimental campaigns, aiming to provide explicit uncertainty bounds for all key metrics.

In summary, the combined AM + SM approach represents a significant advancement in the efficient and reliable fabrication of aeroelastic wind tunnel models. By systematically addressing the limitations of previous AM-based methods, it paves the way for more accurate, reproducible, and cost-effective aeroelastic experimentation.

## 8. Conclusions

This paper presented a new methodology to effectively construct wing models for high-speed wind tunnel testing by combining metal AM and SM processes.

To demonstrate the feasibility and capability of wing models fabricated by the present approach, their geometrical accuracy and structural/aeroelastic characteristics were thoroughly investigated. The current manufacturing method showed a promising improvement in geometrical accuracy and surface roughness compared to previous AM-based methods (e.g., those using MP), providing sufficient quality for aerodynamic performance without undesirable effects. A significant advantage of the present technique is its ability to provide stable quality and high reproducibility for constructing multiple models, achieved by introducing CNC machining and avoiding time-consuming, skill-dependent manual polishing. This approach also proved capable of precisely realizing complex, practical wing designs incorporating sweep, double-taper ratio, and cambered airfoil profiles.

Experimental evaluations confirmed that the final models accurately realized their designed stiffness and modal characteristics. While simpler Wing 1 showed excellent agreement with numerical solutions in both static and modal tests, the more complex Wing 2 exhibited slight discrepancies in higher-mode natural frequencies. This suggests that for highly intricate geometries, further investigation (e.g., using X-ray CT scan for cross-sectional geometry) and optimization of AM and machining parameters are warranted.

Finally, transonic flutter wind tunnel tests with the fabricated Wing 1 models demonstrated that the combined manufacturing technique is feasible for high-speed aeroelastic testing. The excellent reproducibility observed in the flutter boundaries and frequencies between independently manufactured models further validates the robust nature of the proposed approach. This good reproducibility, enabling reliable and repeatable test data, is critical for advancing research in nonlinear aeroelasticity through precise and effective wind tunnel evaluations.

## CRediT authorship contribution statement

**Natsuki Tsushima:** Writing – review & editing, Writing – original draft, Visualization, Validation, Supervision, Software, Resources, Project administration, Methodology, Investigation, Funding acquisition, Data curation, Conceptualization. **Kensuke Soneda:** Writing – review & editing, Investigation, Data curation. **Kenichi Saitoh:** Writing – review & editing, Resources, Investigation. **Kazuyuki Nakakita:** Writing – review & editing, Supervision, Resources, Project administration, Investigation.

## Declaration of competing interest

The authors declare the following financial interests/personal relationships which may be considered as potential competing interests:

Natsuki Tsushima reports financial support was provided by Japan Society for the Promotion of Science. If there are other authors, they declare that they have no known competing financial interests or personal relationships that could have appeared to influence the work reported in this paper.

## Acknowledgements

Part of the research was conducted under the financial support of Grant-in-Aid for Scientific Research (24K17450) by Japan Society for the Promotion of Science.

## Data availability

Data will be made available on request.